\documentclass[useAMS,usenatbib]{mn2e}
\usepackage{url,times,graphicx,amsmath,amsfonts,amssymb,aas_macros,color,epsfig}

\newcommand{\Tab}[1]{Table~\ref{#1}}
\newcommand{\Sec}[1]{Section~\ref{#1}}
\newcommand{\Eq}[1]{Eq.~(\ref{#1})}
\newcommand{\Fig}[1]{Fig.~\ref{#1}}
\newcommand{\beq}{\begin{equation}}
\newcommand{\eeq}{\end{equation}}
\newcommand{\hMpc}{{\ifmmode{h^{-1}{\rm Mpc}}\else{$h^{-1}$Mpc}\fi}}
\newcommand{\hGpc}{{\ifmmode{h^{-1}{\rm Mpc}}\else{$h^{-1}$Gpc}\fi}}
\newcommand{\hkpc}{{\ifmmode{h^{-1}{\rm kpc}}\else{$h^{-1}$kpc}\fi}}
\newcommand{\hMsun}{{\ifmmode{h^{-1}{\rm {M_{\odot}}}}\else{$h^{-1}{\rm{M_{\odot}}}$}\fi}}
\def\hMpc{$h^{-1}\,{\rm Mpc}$}
\def\hkpc{$h^{-1}\,{\rm kpc}$}
\def\LCDM{\ensuremath{\Lambda}CDM}

\title[$N$-body simulations with a cosmic vector for dark energy]
      {$N$-body simulations with a cosmic vector for dark energy}
\author[Carlesi et al.]
       {Edoardo Carlesi,$^{1}$\thanks{E-mail: edoardo.carlesi@uam.es} 
	Alexander Knebe,$^{1}$ 
	Gustavo Yepes,$^{1}$ Stefan Gottl\"ober,$^{3}$ \newauthor
	Jose Beltr\'an Jim\'enez,$^{4,5}$ Antonio L. Maroto,$^{2}$ 
\\
$^{1}$Departamento de F\'isica Te\'orica, Universidad Aut\'onoma de Madrid, 28049, Cantoblanco, Madrid, Spain\\
$^{2}$Departamento de F\'isica Te\'orica, Universidad Complutense de Madrid, 28040, Madrid, Spain \\
$^{3}$Leibniz Institut f\"ur Astrophysik, An der Sternwarte 16, 14482, Potsdam, Germany \\
$^{4}$Institute de Physique Th\'eorique and Center for Astroparticle Physics, Universit\'e de Gen\`eve, 24 quai E. Ansermet, 1211 Gen\`eve, Switzerland\\
$^{5}$Institute of Theoretical Astrophysics, University of Oslo, 0315 Oslo, Norway
}

\begin{document}

\date{Accepted XXXX . Received XXXX; in original form XXXX}

\pagerange{\pageref{firstpage}--\pageref{lastpage}} \pubyear{2011}

\maketitle

\label{firstpage}

\begin{abstract}
We present the results of a series of cosmological $N$-body simulations of a Vector Dark Energy (VDE) model, 
performed using a suitably modified version of the publicly available \texttt{GADGET}-2 code.
The setups of our simulations were calibrated pursuing a twofold aim: 1) to analyze the large scale distribution 
of massive objects and 2) to determine the properties 
of halo structure in this different framework.
We observe that structure formation is enhanced in VDE, since the mass function at high redshift
is boosted up to a factor of ten with respect to \LCDM, 
possibly alleviating tensions with the observations of massive 
clusters at high redshifts and early reionization epoch. 
Significant differences can also be found for the value of the growth factor, that in VDE shows 
a completely different behaviour, and in the distribution of voids, which in this cosmology
are on average smaller and less abundant. 
We further studied the structure of dark matter haloes more massive than $5\times10^{13}$\hMsun, finding 
that no substantial difference emerges when comparing spin parameter, shape, triaxiality and profiles
of structures evolved under different cosmological pictures.
Nevertheless, minor differences can be found in the concentration-mass relation and the two point
correlation function; both showing different amplitudes and steeper slopes.
Using an additional series of simulations of a \LCDM\ scenario with the same $\Omega_M$ and $\sigma_8$ 
used for the VDE cosmology, we have been able to establish wether the modifications
induced in the new cosmological picture were due to the particular nature of the dynamical dark energy
used or a straightforward consequence of the cosmological parameters used.
On large scales, the dynamical effects of the cosmic vector can be seen in 
the peculiar evolution of the cluster number density function with redshift, in the distribution of voids
and on the characteristic form of the growth index $\gamma(z)$. 
On smaller scales, internal properties of haloes are almost unaffected by the change of cosmology, since
no statistical difference can be observed in the characteristics of halo profiles, 
spin parameters, shapes and triaxialities. 
Only halo masses and concentrations show a substantial increase, which can however be attributed to the change 
in the cosmological parameters.
\end{abstract}


\section{Introduction} \label{sec:introduction}
During the last twelve years, a large amount of cosmological high precision data on
Supernovae Ia \citep[see][]{Riess:1999, Perlmutter:1999, SNLS:2010}
cosmic microwave background anisotropies \citep[][]{Wmap:2011, Sherwin:2011}, 
weak lensing \citep{Huterer:2010}, baryon acoustic oscillations 
\citep{Beutler:2011} and large scale structure surveys \citep{SDSS:2009}  
has provided evidence that the Universe we live in is of a flat geometry and undergoing an accelerated expansion.
These observations motivate our belief in the existence 
of an ubiquitous fluid called dark energy (DE) 
that by the exertion of a negative pressure, 
counters and eventually overcomes the gravitational attraction
that would otherwise dominate the evolution of our Universe.
The simplest explanation to the nature of this fluid
is found in the standard model of cosmology 
\LCDM, where the role of the DE is played by a cosmological constant $\Lambda$
obeying the equation of state $p_{\Lambda}=-\rho_{\Lambda}$.
Although perfectly consistent with all the aforementioned observations, \LCDM\ still
lacks of appeal from a purely theoretical point of view.
In fact, if we belive the cosmological constant to be the zero point energy 
of some fundamental quantum field, 
its introduction in the Friedmann equations requires 
a fine-tuning of several tens of orders of magnitude 
(depending on the energy scale we choose to be fundamental in our theory) 
spoiling the naturalness of the whole \LCDM\ picture. \\
Another issue we encounter when dealing with the standard cosmological
model is the so called \emph{coincidence problem}, that is, 
the difficulty to explain in a natural way the fact that 
today's matter and dark energy densities have a comparable value 
although they evolved in a completely different manner throughout 
most of the history of the universe.\\
In an attempt to overcome these two difficulties of \LCDM, 
\cite{Jimenez:2008er} introduced the Vector Dark Energy (VDE) model,
where a cosmic vector field plays the role of a dynamical dark energy
component, replacing the cosmological constant $\Lambda$.
Besides being compatible with supernovae observations and CMB precision measurements, this scenario
has the same number of free parameters as \LCDM. Moreover, the initial value of the vector field 
(which is of the order of $10^{-4}M_p$\footnote{$M_p$ being the Planck mass.}, a scale that could arise 
naturally in inflation)
and its global dynamics ensure the model to overcome the standard model's naturarlness problems.
In the present work we study the impact of this VDE model on structure formation and evolution
by means of a series of cosmological $N$-body simulations, analyzing the effects of this alternative cosmology
in the deeply non-linear regime and highlighting its imprints on cosmic structures, in particular, 
emphasizing the differences emerging with respect to the standard model \LCDM.

The paper is organized as follows. In \Sec{sec:model} we briefly introduce the VDE model,
discussing its most important
mathematical and physical characteristics. In \Sec{sec:setup} we describe the 
setup as well as the modifications to the code and the initial conditions necessary to run the 
$N$-body simulation. In the two Sections~\ref{sec:results1} and \ref{sec:results2}
we will present a detailed analysis of the results, 
focusing on the main differences of the VDE models to the standard \LCDM\ cosmology, first
analyzing the large-scale structure and then (cross-)comparing properties of dark matter haloes.
A short summary of the results obtained and a discussion on their implications is then presented in section 
\Sec{sec:conclusions}.

\section{The Model} \label{sec:model}
\begin{figure*}
\begin{center}
$\begin{array}{cc}
\includegraphics[width=8cm]{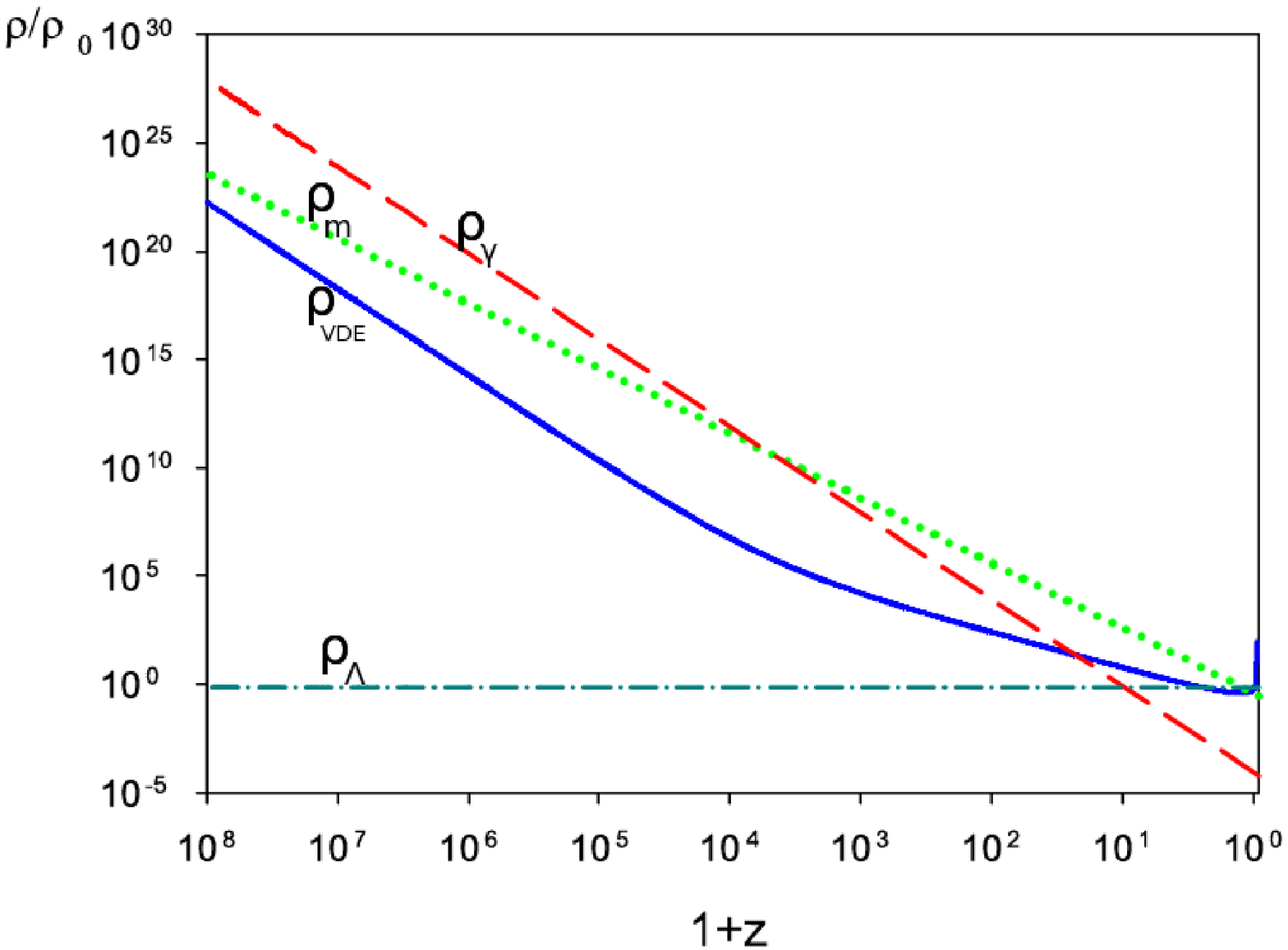}
\includegraphics[width=8.2cm]{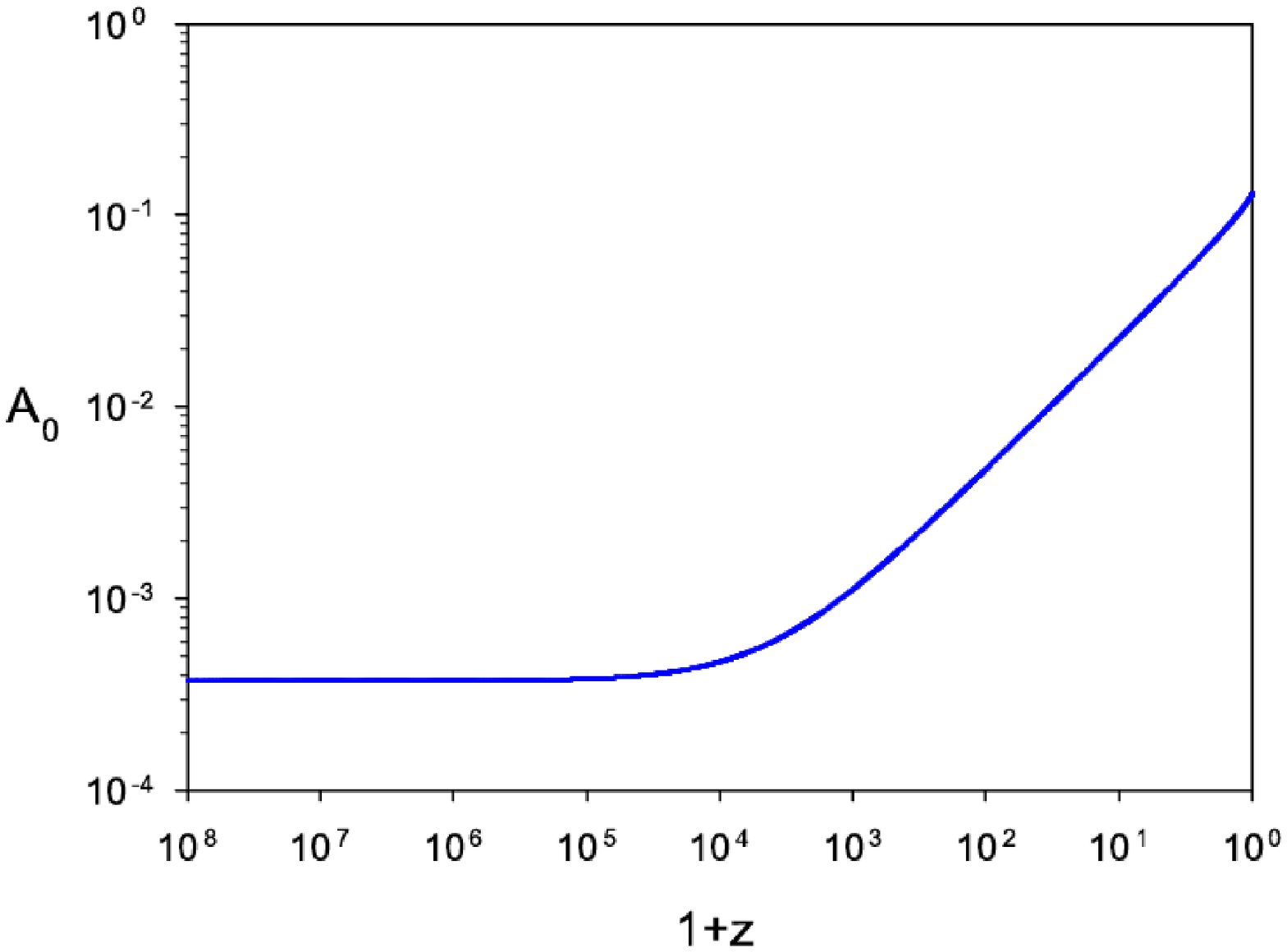}
\end{array}$
\caption{\small
Left plot: Evolution of the energy densities. Dashed (red) for radiation, dotted
(green) for matter and solid (blue) for vector dark energy. We
also show for comparison the cosmological constant energy density
in dashed-dotted line. We see the scaling behaviour of the cosmic vector in the early universe and the rapid growth of its energy density
contribution at late times when approaching the final singularity. Right plot: Evolution of the temporal component of the vector field where we see that it takes a constant value at very high redshifts so that the cosmological evolutions is insensitive to the precise redshift at which we set the initial value of the cosmic vector.
}\label{cosmoevol}
\end{center}
\end{figure*}

In this section we will provide the basic mathematical and physical
description of the VDE model. For more details and an in-depth discussion
on the results obtained and their derivation we refer to \cite{BeltranMaroto:2008}.
The action of the proposed vector dark energy model can be written as:
\begin{eqnarray}
S=\int d^4x \sqrt{-g}\left[-\frac{R}{16\pi G}
-\frac{1}{4}\textbf{\emph{F}}_{\mu\nu}\textbf{\emph{F}}^{\mu\nu}\right. \nonumber\\
\left.-\frac{1}{2}\left(\nabla_\mu
\textbf{\emph{A}}^\mu\right)^2+ \textbf{\emph{R}}_{\mu\nu}\textbf{\emph{A}}^\mu \textbf{\emph{A}}^\nu\right]. \label{CVaction}
\end{eqnarray}
where $\textbf{\emph{R}}_{\mu\nu}$ is the Ricci tensor, $R=\textbf{\emph{g}}^{\mu\nu}\textbf{\emph{R}}_{\mu\nu}$ the scalar curvature and $\textbf{\emph{F}}_{\mu\nu}=\partial_\mu \textbf{\emph{A}}_\nu-\partial_\nu \textbf{\emph{A}}_\mu$. This action can be interpreted as a Maxwell term for a vector field supplemented with a gauge-fixing term and an effective mass provided by the Ricci tensor. It is interesting to note that the vector sector has no free parameters nor potential terms, being $G$ the only dimensional constant of the theory. This is one of the main differences of this model with respect to those based on scalar fields, which need the presence of potential terms to be able to lead to late-time accelerated expansion.

The classical equations of motion derived from the action (\ref{CVaction}) are the Einstein and vector field equations
given by:
\begin{eqnarray}
 \textbf{\emph{R}}_{\mu\nu}-\frac{1}{2}R \textbf{\emph{g}}_{\mu\nu}&=&8\pi G (\textbf{\emph{T}}_{\mu\nu}+\textbf{\emph{T}}_{\mu\nu}^A), \label{eqE}\\
\Box \textbf{\emph{A}}_\mu + \textbf{\emph{R}}_{\mu\nu}\textbf{\emph{A}}^\nu&=&0, \label{eqA}
\end{eqnarray}
where $\textbf{\emph{T}}_{\mu\nu}$ is the conserved energy-momentum tensor for
matter and radiation (and/or other possible components present in the
Universe) and $\textbf{\emph{T}}_{\mu\nu}^\textbf{\emph{A}}$ is the energy-momentum tensor coming
from the vector field sector (and that is also covariantly conserved). In the following, we shall solve the equations of the vector field during the radiation and matter eras, in which the contribution of dark energy is supposed to be negligible. In those epochs, the geometry of the universe is well-described by the flat Friedmann-Lema\^itre-Robertson-Walker metric:
\begin{equation}
ds^2=dt^2-a(t)^2d\textbf{\emph{x}}^2.
\end{equation}
For the homogeneous vector field we shall assume, without lack of generality, the form
$A_\mu=(A_0(t),0,0,A_z(t))$ so that the corresponding equations read:
\begin{eqnarray}
\ddot{A}_0+3H\dot{A}_0-3\left(2H^2+\dot{H}\right)A_0=0,\\ 
\ddot{A}_z+H\dot{A}_z-2\left(\dot{H}+3H^2\right)A_z=0,
\end{eqnarray}
where $H=\dot{a}/a$. These equations can be easily solved for a power law expansion with $H=p/t$, in which case we obtain the following solutions:
\begin{eqnarray}
A_0(t)=A_0^+t^{\alpha_+}+A_0^-t^{\alpha_-}\label{fieldsol0},\\
A_z(t)=A_z^+t^{2p}+A_z^-t^{1-3p}.\label{fieldsol}
\end{eqnarray}
with $\alpha_{\pm}=\frac{1}{2}(1-3p\pm\sqrt{33p^2-18p+1})$ and $A_0^\pm$ and $A_z^\pm$ constants of integration. Thus, in the radiation dominated epoch ($p=1/2$) we have the growing modes $A_0=$constant and $A_z\propto t$, whereas for the matter dominated epoch ($p=2/3$) we have $A_0\propto t^{(-3+\sqrt{33})/6}$ and $A_z\propto t^{4/3}$. Concerning the energy densities, the corresponding expressions are given by:
\begin{eqnarray}
&&\rho_{A_0}=\frac{3}{2}H^2A_0^2+3HA_0\dot A_0-\frac{1}{2}\dot A_0^2,\\
&&\rho_{A_z}=\frac{1}{2a^2}\left(4H^2A_z^2-4HA_z\dot{A}_z+\dot{A}_z^2\right).
\end{eqnarray}
At this point, it is interesting to note that when we insert the full solution for $A_z$ given in (\ref{fieldsol}) in its corresponding energy density, we obtain:
\begin{eqnarray}
\rho_{A_z}=\frac{\left(A_z^-\right)^2}{2a^8}\left(25p^2-10p+1\right).
\end{eqnarray}
so that the mode $A_z^+$ does not contribute to the energy density. That way, even though $A_z$ grows with respect to $A_0$, the corresponding physical quantity, i.e. its energy density, decays with respect to that of the temporal component. It is easy to check that the ratio $\rho_{A_z}/\rho_{A_0}$ decays as $a^{-4}$ in the radiation era and as $a^{-6.37}$ in the matter era so that the energy density of the vector field becomes dominated by the contribution of the temporal component. That justifies to neglect the spatial components and deal uniquely with the temporal one. 

On the other hand, the potential large scale anisotropy generated by the presence of spatial components of the vector field is determined by the relative difference of pressures in different directions $p_\parallel$ and $p_\perp$, that is given by:
\begin{eqnarray}
p_\parallel-p_\perp=
\frac{3}{a^2}\left(4H^2A_z^2-4HA_z\dot{A}_z+\dot{A}_z^2\right).
\end{eqnarray}
This expression happens to be proportional to $\rho_{A_z}$ so that we have that $(p_\parallel-p_\perp)/\rho_A$ will decay as the universe expands in the same manner as $\rho_{A_z}$ and the large scale isotropy of the universe suggested by CMB observations is not spoiled. Hence, in the following we shall neglect the spatial components of the vector field and we shall uniquely consider the temporal one, since it gives the dominant contribution to the energy-momentum tensor of the vector field. However, we should emphasize here that this does not result in effectively having a scalar field. As commented before, for a minimally-coupled scalar field, one needs to introduce a certain potential (that will depend on some dimensional parameters) to have accelerated expansion, whereas in the VDE model, we get accelerated solutions with only kinetic terms and without introducing any new dimensionful parameter.

The energy density of the vector field is given by:
\begin{eqnarray}
\rho_{A}= \rho_{A0} (1+z)^\kappa,
\end{eqnarray}
with $\kappa=4$ in the radiation era and $\kappa=(9-\sqrt{33})/2
\simeq -1.63$ in the matter era.  We can also calculate the effective equation of state for dark energy as:
\begin{eqnarray}
w_{DE}=\frac{p_A}{\rho_A}=\frac{-3\left(\frac{5}{2}H^2+\frac{4}{3}\dot{H}\right)A_0^2+HA_0\dot{A}_0
-\frac{3}{2}\dot{A}_0^2}{\frac{3}{2}H^2A_0^2+3HA_0\dot
A_0-\frac{1}{2}\dot A_0^2}.
\end{eqnarray}
Again, using the approximate solutions in (\ref{fieldsol0}), we
obtain;
\begin{eqnarray}
w_{DE}=\left \{
\begin{array}{l}
\frac{1}{3}\;\;\;\;\;\;\;\;\;\;\;\;\;\;\;\;\;\;\;\;\;\;\;\;\;\;\;\;\;\; \mbox{radiation era}\\
\frac{3\sqrt{33}-13}{\sqrt{33}-15}\simeq
-0.457\;\;\;\;\;\mbox{matter era}
\end{array}
\right.
\end{eqnarray}
From the evolution of the energy density of the
vector field we see that it scales as radiation at early times, so that $\rho_A/\rho_R=$ const. However, when the
Universe enters its matter era, $\rho_A$ starts growing relative
to $\rho_M$ eventually overcoming it at some point, in which the
dark energy vector field would become the dominant component. From
that point on,  we cannot obtain analytic solutions to the field
equations and we need to numerically solve the corresponding equations. In Fig. \ref{cosmoevol} we show such a numerical solution
to the exact equations, which confirms our analytical estimates in
the radiation and matter eras.  Notice that, since $A_0$ is constant during the radiation era, the solutions do not
depend on the precise time at which we specify the initial
conditions as long as we set them well inside the radiation epoch.
Thus, once the present value of the Hubble parameter $H_0$ and the
constant $A_0$ during radiation (which indirectly fixes the total matter
density $\Omega_M$) are specified, the model is completely
determined. In other words, this model contains the same number of
parameters as $\Lambda$CDM, i.e. the minimum number of parameters
of any cosmological model with dark energy. 

The vector dark energy model does not only have the minimum required number of 
parameters, but it also allows to alleviate the so-called naturalness or 
coincidence problems that most dark energy models have. This is so because 
the required value for the constant value that the vector field takes in the 
early universe happens to be $\sim 10^{-4} M_p$. This value, in addition to 
be relatively close to the Planck scale, could naturally arise from quantum 
fluctuations during inflation, for instance. On the other hand, the fact 
that the energy density of the vector field scales as radiation in the early 
universe also goes in the right direction of alleviating the aforementioned 
problems because the fraction of dark energy during that period remains constant. 
Moreover, said fraction is $\Omega_A^{early}\equiv\rho_A/\rho_R\simeq 10^{-6}$, 
which, again, is in agreement with the usual magnitude of the quantum fluctuations 
produced during inflation.

\begin{figure}
\centering
\includegraphics[width=8cm]{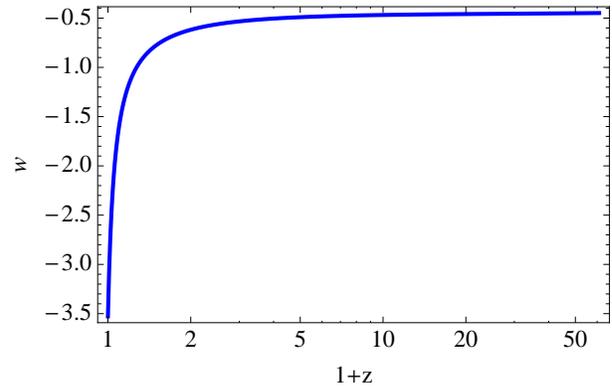}
\caption{\small Evolution of dark energy equation of
state where we can see the crossing of the phantom divide line and its evolution towards $-\infty$ as the final singularity is approached. }\label{EoS}
\end{figure}

After dark energy starts dominating, the equation of state
abruptly falls towards $w_{DE}\rightarrow -\infty$ as the Universe
approaches a finite time $t_{end}$. As shown in Fig. \ref{EoS}, during the cosmological evolution the equation
of state crosses the so-called phantom divide line, so that we
have $w_{DE}(z=0)<-1$. The final stage of the universe in this model is a 
singularity usually called Type III or Big Freeze, in which the scale factor 
remains finite, but the Hubble expansion rate, the energy density and the pressure diverge. This is a distinct feature of the VDE model as compared to quintessence fields for which the equation of state is restricted to be $>-1$ so that no crossing of the phantom divide line is possible. In fact, for a dark energy model based on scalar fields, one needs either non-standard kinetic terms involving higher derivative terms in the action or the presence of several interacting scalar fields to achieve a transition from $w>-1$ to a phantom behaviour ($w<-1$). In either case, non-linear derivative interactions or multiple scalar field scenarios, additional degrees of freedom are introduced, whereas the VDE model is able to obtained the mentioned transition with only the degree of freedom given by the temporal component of the vector field.

Notice that in the VDE model
the present value of the equation of state parameter $w_0 = -3.53$ is
radically different from that of a cosmological constant (cf. Fig. 1,
where the redshift evolution of $w(z)$ is shown in the range of our
simulations). The values of other cosmological parameters 
also differ importantly from those of $\Lambda$CDM (see \Tab{tab:settings}). Despite this fact, 
VDE is able to simultaneously fit
supernovae and CMB data with comparable goodness to $\Lambda$CDM
(\cite{BeltranMaroto:2008}, \cite{BLM:2009}). This might seem to be surprising if we notice that the present equation of state for the VDE model is $w_0 = -3.53$, which is far from the usual constraints on this parameter obtained from cosmological observations. Such constraints are usually obtained by assuming a certain parametrization for the time variation of the dark energy equation of state. However, the different used parameterizations are normally such that $\Lambda$CDM is included in the parameter space. If we look at Fig. \ref{EoS}, we can see that the evolution of the equation of state for the VDE model crucially differs from those of $\Lambda$CDM or quintessence models and, indeed, it cannot be properly described by the most popular parameterizations. This means that we cannot directly apply the existing constraints to the VDE model, but a direct comparison of its predictions to observations is required.

As a final remark, in the simulations we will not include inhomogeneous perturbations of the vector field, but only the effects of having a different background expansion will be considered. 

In Fig. 3, we show the matter power spectrum for both $\Lambda$CDM and the VDE model. The differences can be ascribed to the fact of having different cosmological parameters that change the normalization and the matter-radiation equality scale $k_{eq}$, which are the only two differences observed. Notice that the transfer function is the same in both cases, since the slopes before and after the $k_{eq}$ are the same, so that we do not expect strong effects at early times which could affect the evolution of density parameters.

In particular, for CMB\footnote{We use the binned data of WMAP7}, the $\chi^2$ for the best fit parameters for $\Lambda$CDM is $48.3$, wheres for the VDE model we obtain $\chi^2=51.8$ for the parameters used to run our simulations. Thus, even though the equation of state evolution is as the one shown in Fig. \ref{EoS}, the VDE model provides good fits to observations.

\section{The $N$-Body Simulations} \label{sec:setup}

In this section we will explain the (numerical) methods used in this work, with a particular
emphasis on the necessary modifications of the standard $N$-body and halo finding algorithms, also describing 
the procedures followed to test their accuracy and reliability.

\subsection{Set-up}
The $N$-body simulations presented in this work have been carried out using a suitably modified
version of the Tree-PM code \texttt{GADGET}-2 \citep{Springel:2005}. 
It has been also necessary to generate a particular 
set of initial conditions to consistently account for the VDE-induced
modifications to the standard paradigm.
In \Tab{tab:settings} we show the most relevant cosmological parameters used in
the different simulations. 
For the VDE model, we have used the value of $\Omega_M$ 
provided by the best fit to SNIa data, whereas the remaining cosmological parameters have
been obtained by a fit of the model to the WMAP7 dataset. 
For \LCDM\ we used the
Multidark Simulation \citep{Prada:2011} cosmological parameters
with a WMAP7 $\sigma_8$ normalization \citep{Wmap:2011}.

In addition, we also simulated a so-called \LCDM-vde model, which implements
the VDE values for the total matter density $\Omega_M$ and fluctuation amplitude $\sigma_8$ in an otherwise standard \LCDM\ picture.
Although ruled out by current cosmological constraints, 
this model provides nonetheless an interesting case study that allows us 
to shed light on the effects of these two cosmological parameters on 
structure formation in the VDE model. 
In particular, we want to be able to determine the impact of the different parameters 
on cosmological scales, with a particular emphasis on the very large structures and 
the most massive clusters, where observations are starting to clash with the predictions
of the current standard model (see \cite{Jee:2009}, \cite{Baldi:2010}, 
\cite{Hoyle:2010}, \cite{Carlesi:2011} and \cite{Enqvist:2011}). 
Therefore, we need to determine whether the results derived from our VDE simulations can be solely attributed 
to its extremely different values for the cosmological parameters or actually by the presence of the cosmic vector field. 
In other words, we want to separate the signatures of the \emph{dynamics}-driven effects from the 
\emph{parameter}-driven ones, with a focus on large scale structures, where the imprints 
are stronger and more clearly connected to the cosmological model.
We chose to run a total of eight $512^3$ particle simulations
summarized in \Tab{tab:settings} and explained below:

\begin{itemize}
\item two VDE simulations, i.e. a 500 \hMpc\ and a 1 \hGpc\ box,
\item two \LCDM\ simulations with the same box sizes and initial seeds as the VDE runs above,
\item two more VDE simulations with a different random seed, again one in a 500 \hMpc$\;$
and one in a 1 \hGpc$\;$box (both serving as a check for the influence of cosmic variance), and
\item two \LCDM-vde simulations, one again in a 500 \hMpc\ and one in a 1 \hGpc\ box.
\end{itemize}

All runs were performed on 64 CPUs using the MareNostrum cluster at the 
Barcelona Supercomputing Center.
Most of the results we will discuss and analyze here are based on the 500 \hMpc\ simulations as they have the better mass resolution. The
1~\hGpc\ runs primarily serve as a confirmation of the results and have already been discussed in \cite{Carlesi:2011}, respectively.

\begin{table}
\caption{$N$-body settings and cosmological parameters used for the \texttt{GADGET}-2 simulations, 
  the two 500\hMpc\ and the two 1\hGpc\ have the same initial random seed (in order to allow for a direct comparison of the halo
  properties) and starting redshift $z_{\rm start}=60$.
  The number of particles in each run was fixed at $512^3$.
The box size $B$ is given in units of \hMpc\ and the particle
  mass in $10^{11}$ \hMsun.}
\begin{center}
\begin{tabular}{lcccccc}
\hline
Simulation &  $\Omega_{m}$ & $\Omega_{de}$ & $\sigma_8$ & h & $B$ & $m_{p}$ \\
\hline
$2\times$VDE-0.5 & 0.388 & 0.612 & 0.83 & 0.62 &  500  & $1.00$ \\
$2\times$VDE-1   & 0.388 & 0.612 & 0.83 & 0.62 &  1000 & $8.02$ \\
$\Lambda$CDM-0.5 & 0.27  & 0.73  & 0.8  & 0.7  &  500  & $0.69$ \\
$\Lambda$CDM-1   & 0.27  & 0.73  & 0.8  & 0.7  &  1000 &  $5.55$ \\
$\Lambda$CDM-0.5vde & 0.388 & 0.612 & 0.83 & 0.7 &  500 & $1.00$ \\
$\Lambda$CDM-1vde   & 0.388 & 0.612 & 0.83 & 0.7 & 1000 & $8.02$ \\
\hline
\end{tabular}
\end{center}
\label{tab:settings}
\end{table}

\subsection{Code Modifications}\label{subsec:modifications}
In the following paragraph we are going to describe the procedures followed to 
implement the modifications needed in order to run our $N$-body simulations
consistently and reliably. 
This is in principle a non-trivial issue, since, as described in \Sec{sec:model}, we need 
to incorporate a large number of different features that affect both the code used
for the simulations and the initial conditions.

\begin{figure}
\centering
\includegraphics[angle=270,width=8cm]{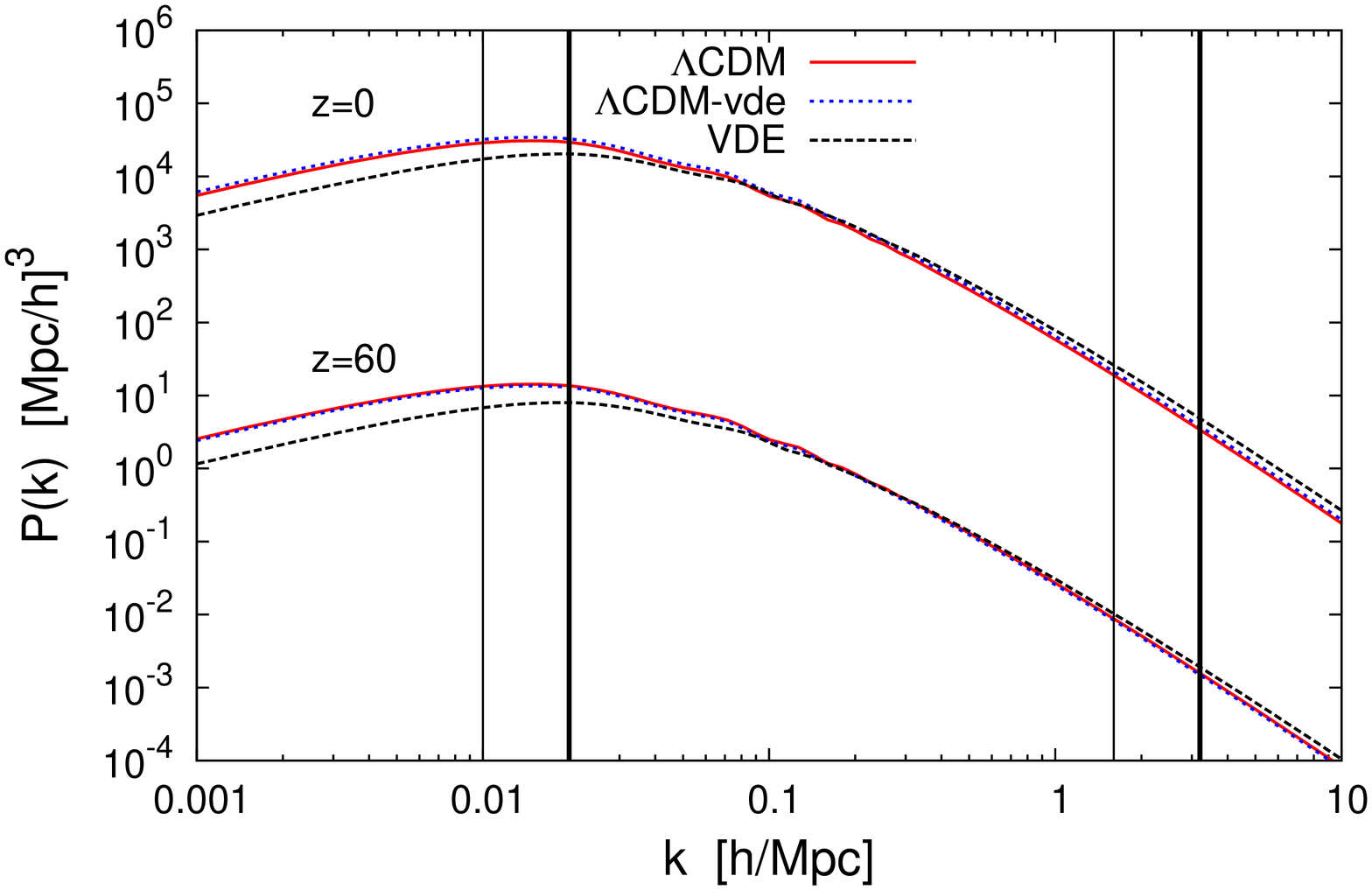}
\caption{\small 
Linear matter overdensity power spectra at $z=0$ and $z=60$ for VDE, \LCDM\ and \LCDM-vde plotted versus
wavenumber $k$. Vertical solid thick black lines refer to the $k$-space interval 
covered by the 500 \hMpc\ simulations whereas the thin ones refer to 
the 1 \hGpc\ one.
All matter power spectra at $z=0$ have been normalized to the $\sigma_8$ values 
shown in \Tab{tab:settings} and then rescaled to $z=60$ via the linear growth factor.
We notice that for $k<0.05$ h Mpc$^{-1}$, \LCDM\ and \LCDM-vde have more power than VDE, whereas on 
smaller scales the opposite is true.
We also note that due to the different value of $\sigma_8$ normalization the \LCDM-vde $P(k)$
is slightly larger than the \LCDM\ one at $z=0$ while the different growth factor, which is larger in the 
\LCDM-vde cosmology, affect the setting of the initial conditions, where the latter power spectrum
lies below the former.
}\label{img:pkz0}
\end{figure}

\begin{figure}
\centering
\includegraphics[angle=270,width=8cm]{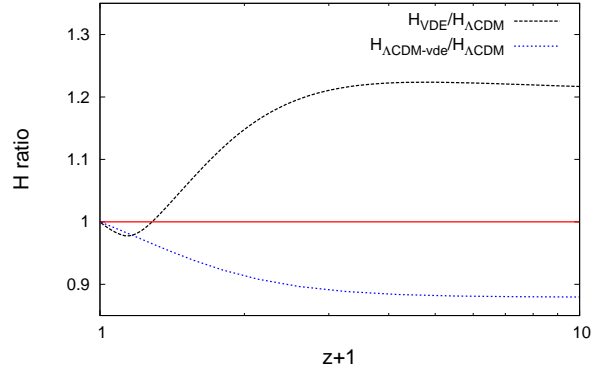}
\caption{\small 
The ratio of the Hubble function $H(a) h^{-1}$ for VDE and
\LCDM-vde to the standard \LCDM\ one.
At earlier times VDE undergoes a relatively faster expansion 
compared to \LCDM, whereas the opposite is true at smaller $z$s.
On the other hand, \LCDM-vde cosmology is characterized by a 
slower relative expansion throughout the whole history of the universe.
}\label{img:hratio}
\end{figure}

In particular, we have
to handle with care three features that distinguish it from \LCDM, i.e.:

\begin{itemize}
\item the matter power spectrum $P(k,z)$ (shown in \Fig{img:pkz0}) and its normalization $\sigma_8$,\\
\item the expansion history $H(z)$ (see \Fig{img:hratio}), and \\
\item the linear growth factor $D^+(z)$ (cf. \Fig{img:growthfactor}).\\
\end{itemize} 

Whereas the first and last point affect the system's initial conditions, 
the second one enters directly into the
$N$-body time integration, and has to be taken into account by a  
modification of the simulation code.

\subsubsection{Initial conditions}
To consistently generate the initial conditions for our simulation, first 
we  normalized the perturbation power spectrum depicted in \Fig{img:pkz0} 
to the chosen value for $\sigma_8$ at $z=0$.
Therefore, we normalized VDE and \LCDM-vde inital conditions
to $\sigma_8=0.83$ while for \LCDM\ we used the WMAP7 value $\sigma_8=0.8$.
Using the respective linear growth factors, we rescaled the $P(k)$ to the initial redshift 
$z=60$ where then the particles' initial velocities and positions were 
computed using the Zel'Dovich \citep{Zeldovich:1970} approximation.
\\
We emphasize here that the main goal of our analysis is to find and highlight 
the main differences of the VDE picture with respect to the standard one: therefore,
the choice of these different normalization parameters has to be understood 
as unavoidable as long as we want the models under investigation to be WMAP7 viable ones.
Needless to say, in this regard the \LCDM-vde cosmology must be considered only as a tool to disentangle
parameter-driven effects from the dynamical ones, not being a concurrent cosmological paradigm
we want to compare VDE to.

\subsubsection{Hubble expansion}
As pointed out by \cite{Li:2011}, the expansion history of the universe has a very
deep impact on structure formation and in particular the results of an $N$-body simulation, as it affects directly every
single particle through the equations of motion written in comoving coordinates.
In \Fig{img:hratio} the ratios of the Hubble expansion factors for VDE and \LCDM-vde to
the standard \LCDM\ value are shown; we see that different models are characterized 
by differences up to the $20\%$ in the expansion rate.
To implement this modification, we  
replaced the standard computation of $H(a)$ in \texttt{GADGET-2} with a routine 
that reads and interpolates from a pre-computed table.

\subsection{Code Testing}
To check the reliability of the modifications introduced into the simulation code and during the generation of the initial conditions, we have 
confronted the theoretical linear growth factor, computed using the 
Boltzmann-code \texttt{CAMB} \citep{Lewis:1999bs} with the ones derived directly 
from the simulations. 

\begin{figure}
\begin{center}
\includegraphics[angle=270, width=8cm]{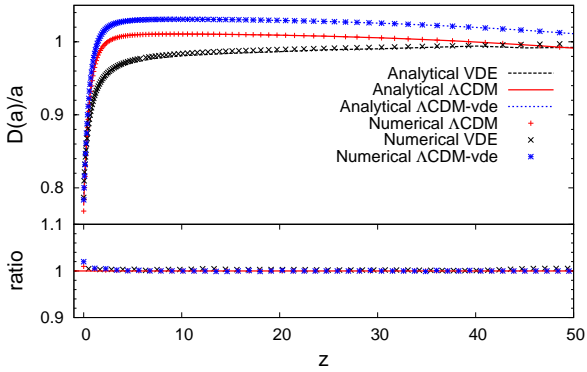}
\caption{Ratio of the growth function to the expansion factor 
$D(a)/a$ as obtained from the 500 \hMpc\ box simulations 
versus the analytical one. 
The results shown an agreement between the theoretical expectation
and the numerically computed one within the $2\%$ level.
The results from the 1 \hGpc\ box simulations are 
not shown since they perfectly overlap with the ones presented here.}
\label{img:growthfactor}
\end{center}
\end{figure}

As shown in \Fig{img:growthfactor}, our results yield an agreement within 
the $1\%$ level, which proofs the correctness of our modifications as well as
illustrating (again) the differences in structure growth between the models.

We would like to note that for consistency reasons, both when calculating the \texttt{CAMB} and the numerical value for the growth factor, 
we have used the expression
\begin{equation}
D^{+}(z) = \sqrt{\frac{P(z,k_0)}{P(z_0,k_0)}}
\end{equation}
where $k_0$ is a fixed scale whithin the linear regime and $z_0$ is the initial redshift of the 
simulation.

\subsection{Halo Finding}\label{sec:analysis}
In order to identify haloes in our simulation we have run the open source
MPI+OpenMP hybrid halo finder \texttt{AHF}\footnote{\texttt{AHF} stands for \texttt{A}miga~\texttt{H}alo~\texttt{F}inder, to be downloaded freely from
\texttt{http://www.popia.ft.uam.es/AMIGA}} described in detail in
\cite{Knollmann:2009}. \texttt{AHF} is an improvement of the
\texttt{MHF} halo finder \citep{Gill:2004} and has been extensively compared against practically all other halo finding methods in \citet{Knebe:2011}.
\texttt{AHF} locates local
overdensities in an adaptively smoothed density field as prospective
halo centres. For each of
these density peaks  the gravitationally bound particles are
determined. Only peaks with at least 20 bound particles are considered
as haloes and retained for further analysis.

But the determination of the mass requires a bit more elaboration as it is computed via the equation

\begin{equation} \label{eq:virial_mass_definition}
M(R) = \Delta \times \rho_{c}(z) \times \frac{4 \pi}{3} R^{3}
\end{equation}
where we applied $\Delta=200$ as the overdensity threshold
and $\rho_c(z)$ refers to the critical density of the universe at redshift $z$. 
In this way $M(R)$ is defined as the total mass contained within a radius $R$, 
corresponding to the point where the halo matter density $\rho(r)$ is $\Delta$ 
times the critical value $\rho_c$.
Using this relation, particular care has to be taken when considering the 
definition of the critical density 
 
\begin{equation}
\rho_{c}(z) = \frac{3 H^2(z)}{8 \pi G} 
\end{equation}
because it involves the Hubble parameter, that differs substantially
at all redshifts in the two models.  This means that, identifying the
halo masses, we have to take into account the fact that the value of
$\rho_c(z)$ changes from \LCDM\ to VDE.  This has been incorporated
into and taken care of in the latest version of \texttt{AHF} where
$H_{VDE}(z)$ is being read in from a precomputed table, too.

We finally need to mention that we checked that the objects obtained by
this (virial) definition can be compared across different cosmological models and using different mass definitions. To this extent we
studied the ratio between two times kinetic over potential energy
$\eta=2T/|U|$ confirming that at each redshift under investigation here
the distributions of $\eta$ in \LCDM\ and VDE are actually comparable (not presented here though), meaning that
the degree of virialisation (which should be guaranteed by \Eq{eq:virial_mass_definition}) is in fact similar. We
therefore conclude that our adopted method to define halo mass (and edge) in the
VDE model leads to unbiased results and yields objects in the same state of equilibrium as is the case for the \LCDM\ haloes. Please note that this test does not guarantee that all our objects are in fact virialized; it merely assures us that the degree of virialisation is equivalent. We will come back to this issue later when selecting only equilibrated objects.


\begin{figure*}
\begin{center}
\includegraphics[width=18cm]{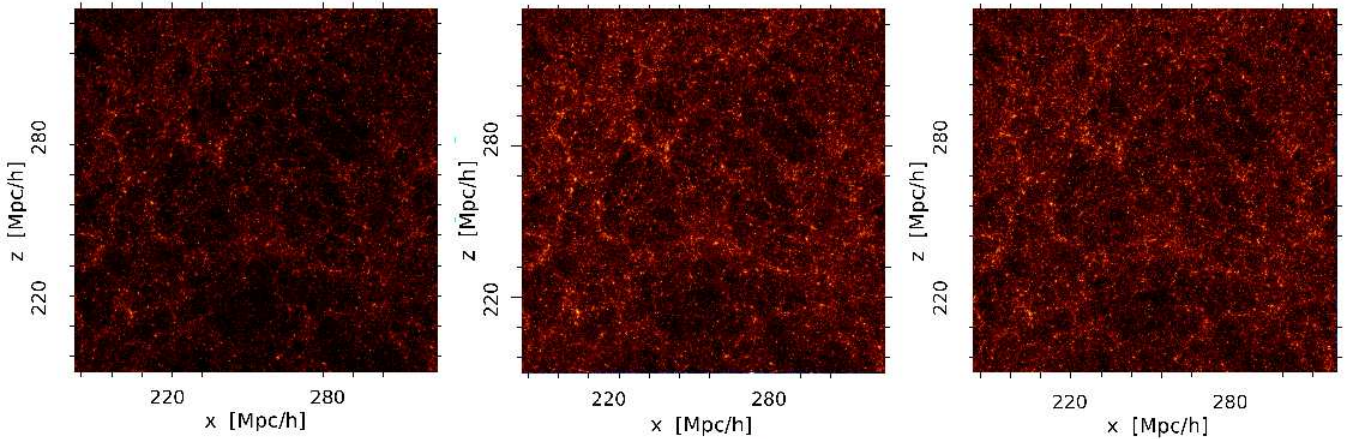}
\caption{Projected density for \LCDM,\LCDM-vde and VDE showing a $120\times120$ $h^{-2}Mpc^2$\ 
slice at the box center in the 500 \hMpc\ box at $z=0$ projected on the $x$-$z$ plane.
Bright areas are associated with matter whereas underdense regions are denoted by 
darker, black spots in the projected box. 
Results for the VDE-1, \LCDM-1 and \LCDM-vde-1 simulations are not shown since the colour coding does not
provide useful insights on the different clustering patterns on smaller scales.
}
\label{img:fullbox}
\end{center}
\end{figure*}

\section{Large Scale Structure and Global Properties} \label{sec:results1}
In the following section, we will discuss the global properties
of large scale structures identified in our simulations.
Using all of our sets of simulations for \LCDM, \LCDM-vde and VDE
we will disentangle parameter-driven effects from those due to the
different dynamics of the background expansion, which uniquely characterize
VDE and therefore are worth pointing out in the process of model selection.

\begin{figure*} 
\begin{center}
\includegraphics[angle=270, width=15cm]{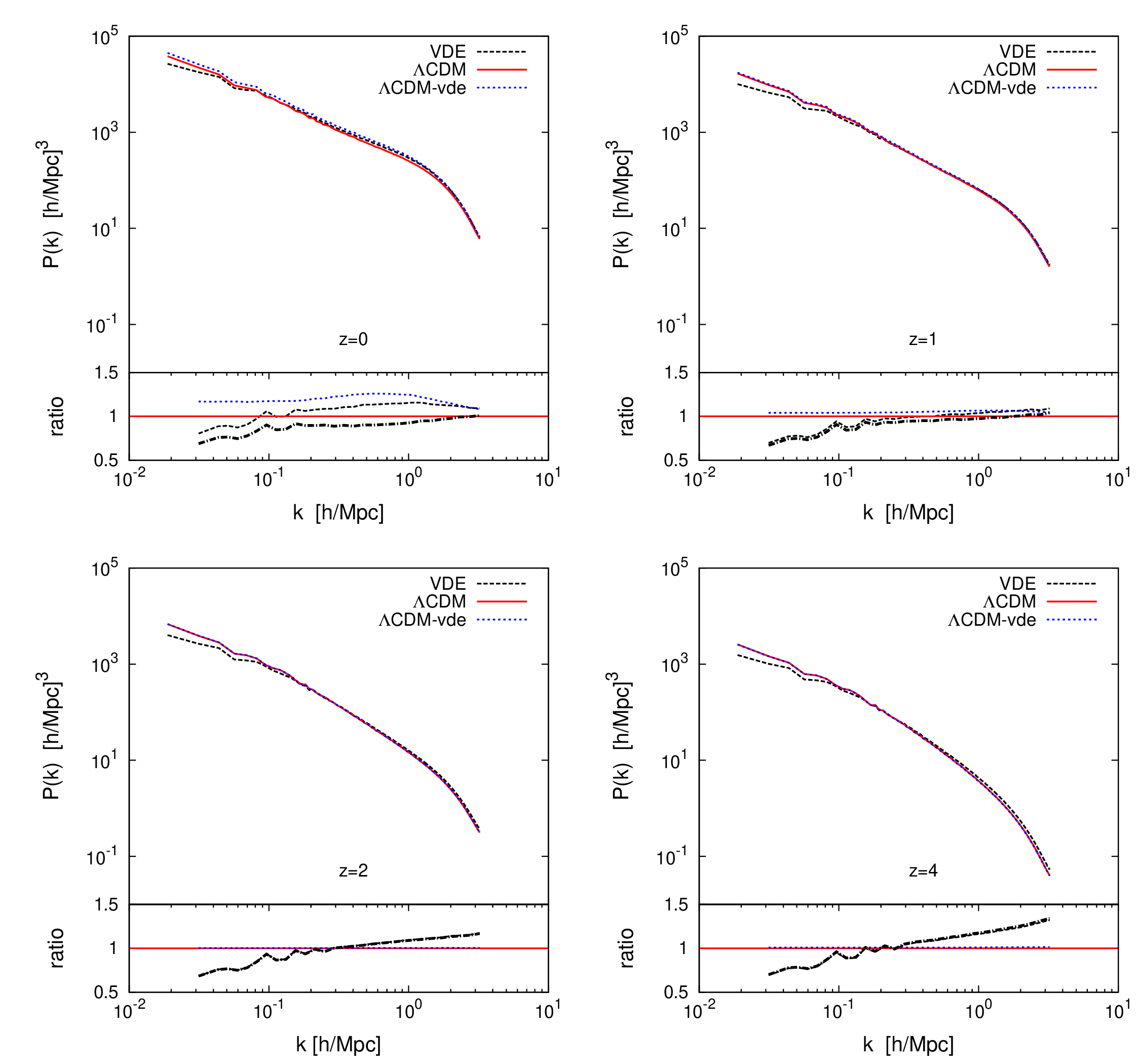}
\caption{Power Spectra at redshifts $z=0,1,2,4$ for \LCDM-0.5, \LCDM-0.5-vde and VDE-0.5
The results from the 1 \hGpc\ simulations are not shown as they simply
overlap to the present ones on the smaller-$k$ end, without providing further insights on the small scales, 
where we expect non linear effects to dominate.
}
\label{img:powerspectra}
\end{center}
\end{figure*}

\begin{figure*}\begin{center}
\includegraphics[angle=270, width=15cm]{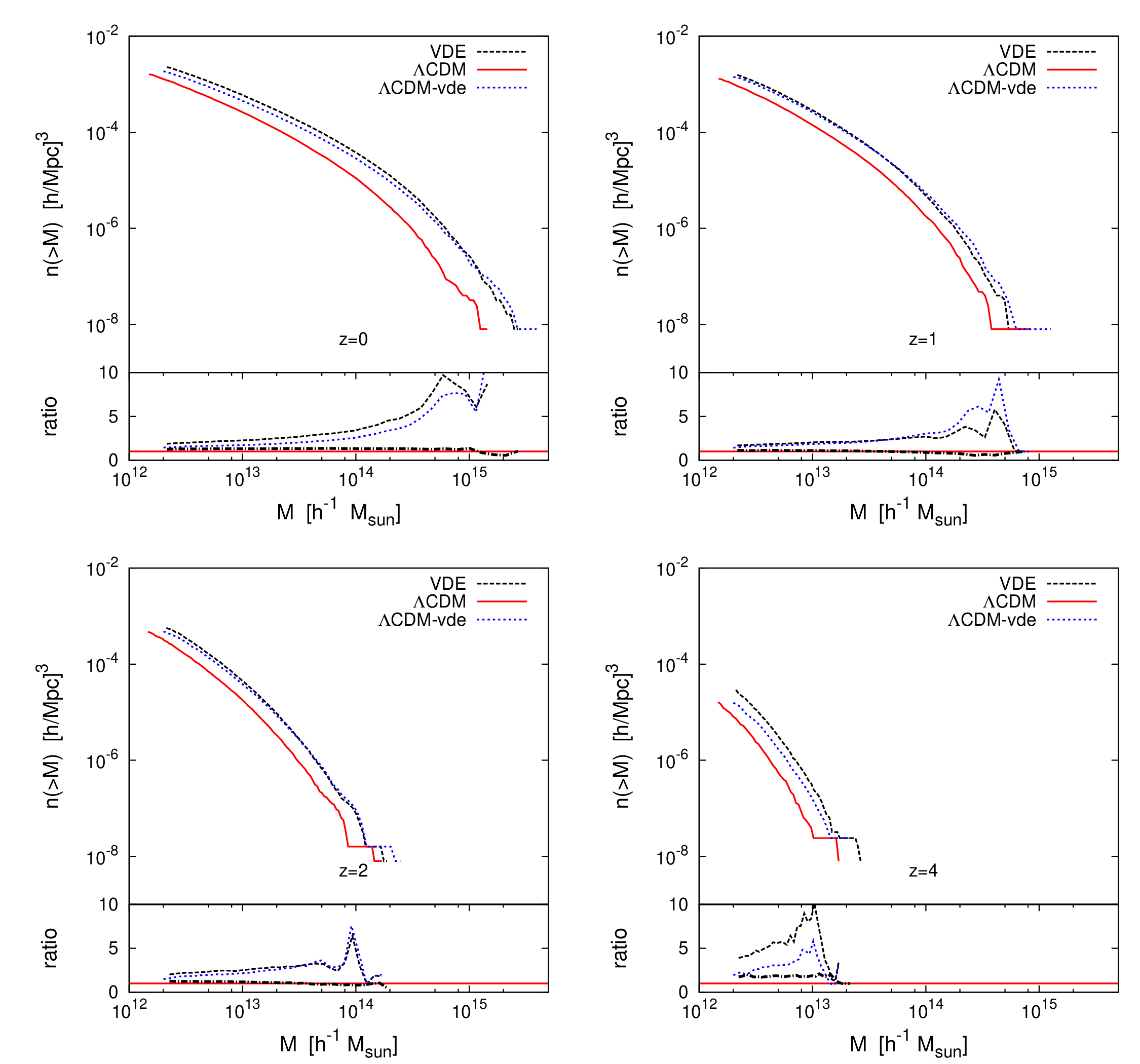}
\caption{Mass function for \LCDM, VDE and \LCDM-vde models at different redshifts, 
computed for the 500 \hMpc\ box simulations. We have also verified that the corresponding values computed for 
the 1 \hGpc\ simulations overlap to the ones shown here for $M>10^{13}$\hMsun, except for a smoother high-mass end.
In the lower panels of the plots, VDE and \LCDM-vde to \LCDM\ ratios are represented by dotted lines 
while VDE to \LCDM-vde are shown using dash-dotted lines.
}
\label{img:massfunction}
\end{center}
\end{figure*} 


\subsection{Density Distribution}
In \Fig{img:fullbox} we show the colour coded 
density field for the particle distribution at redshift $z=0$, 
for a $120\times120$ \hMpc$^2$ slice at the box center
for the three 500 \hMpc\ simulations projected on the $x$-$z$ plane.
As expected, we observe that the most massive structures' spatial positions match 
in the three simulations, although in the \LCDM-vde and VDE a large overabundance
of objects with respect to \LCDM, as we could expect due to the higher $\Omega_M$.
This observation will be confirmed on more quantitative grounds
in the analysis carried in the following sections, especially when refering to the study of 
the cumulative mass function.

\subsection{Matter Power Spectrum} \label{sec:pk}
In \Fig{img:powerspectra} we show the dark matter power spectrum $P(k)$ at redshifts $z=0,1,3,4$ computed 
for the VDE-0.5, \LCDM-0.5 and \LCDM-vde-0.5 simulations. For clarity, 
we do not show  the 1~\hGpc\ simulations; however, we have checked their
consistency with the 500 \hMpc\ runs.
We note that at all redshifts the differences already seen in the input power spectra are preserved (cf. \Fig{img:pkz0}), meaning that
the VDE model has less power 
than \LCDM\ on the large scales, whereas the opposite is true for small scale.
This particular shape of the $P(k)$ is a peculiar feature of VDE cosmology, as other 
kinds of dynamical quintessence \citep{Alimi:2010} and coupled DE \citep{Baldi:2010a} show completely
different properties; with less power (in the former case) or a \LCDM-type of behaviour (in the latter)
on small scales.
At higher and intermediate redshifts \LCDM-vde shows almost no differences from \LCDM, as expected since 
the former is normalized to a lower initial value with respect to the latter and therefore needs to 
equal it before eventually overcoming it at smaller $z$'s, as imposed by the larger $\sigma_8$ normalization.
The effects of the different growth factor in this
model start to become evident only at $z<1$, where we see that the ratio of the $P(k)$ starts to increase.
Whereas the ratio of VDE to \LCDM\ for $k<0.05$ h \hMpc$^{-1}$ is substantially unaltered at all redshifts, 
 small scales are affected by non-linear effects,  eventually distorting its shape. 

\subsection{Halo Abundance}\label{sec:haloab}
In the following subsection we will study the abundance of massive objects at different redshifts.
Highlighting the differences arising among the three models in the different mass
ranges, we want to study VDE's peculiar predictions for the massive cluster distribution 
and highlight its distinction from \LCDM.

To this extent, we show in \Fig{img:massfunction} the (cumulative) mass functions for the three models at $z=0,1,2,4$, as 
computed from the VDE-0.5, \LCDM-0.5 and \LCDM-0.5-vde simulations; 
the corresponding VDE-1, \LCDM-1 and \LCDM-vde-1 results can be found in 
\cite{Carlesi:2011}; they are not shown here again as they do not provide any new insights and rather confirm (and extend) the results to be drawn from the $500$\hMpc\ boxes, respectively: We note that the VDE cosmology is characterized by a larger number 
of objects at all the mass scales and redshifts, 
outnumbering \LCDM\ by a factor constantly larger than 2. In particular, this enhancement
can be seen for the very large masses, where at low $z$ the VDE/\LCDM\ ratio reaches values
of $\sim10$. Although this value of the ratio seems to be a mere result of the cosmic variance, 
due to the low number of haloes found in this mass range,
the computation of the mass function for the second 500 \hMpc\ VDE realization and
the 1\hGpc\ simulations makes us believe that the expected enhancement in this region
must be at least a factor 5.

Interestingly enough, \LCDM-vde has comparable characteristics to VDE, 
which leads to the conclusion that the substantial enhancement in structure 
formation is mainly parameter-driven, i.e.
due to the overabundance of matter and higher normalisation of matter density perturbations.
Although this first observation may seem in contrast with what we have 
found in \Sec{sec:pk}, where we have noticed that
VDE has less power on large scales in comparison to \LCDM, 
we have to take into account that, in the 
hierarchical picture of structure formation, 
objects on small scales form first to subsequently give 
birth to larger ones. 
This means, in our case, that more power for large $k$-values should be regarded 
as an important source of the overall enhancement together 
with the overabundance of matter, as already pointed out in the previous discussion.
The evolution of the mass functions at different redshift allows us to disentangle 
the effect of the 
modified expansion rate; at higher redshift, in fact, 
both the \LCDM\ and \LCDM-vde mass functions are suppressed with respect to the VDE model, 
mostly because of the lack of power on small scales. 
These stronger initial fluctuations eventually trigger 
the earlier start of structure formation, but --
as time passes -- the effect of the increased expansion rate shown in \Fig{img:hratio} 
for the VDE cosmology suppresses structure growth, leading to a mass function below the \LCDM-vde 
curve at around redshift one. At this point, 
the VDE expansion rate starts decreasing with respect to the \LCDM\ one, 
comparatively enhancing very large structure growth and
eventually causing 
the two mass functions to be (nearly) indistinguishable at $z=0$.

\begin{figure}\begin{center} 
\includegraphics[angle=270, width=8cm]{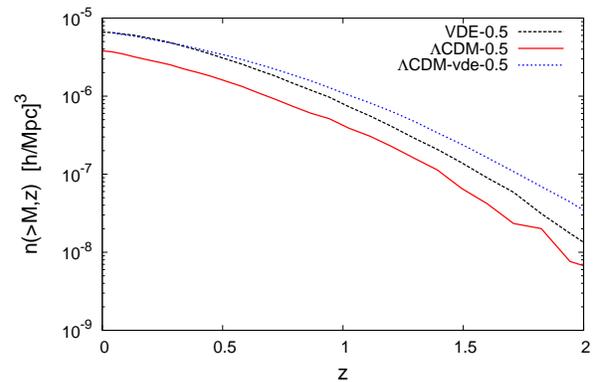}
\caption{Number density evolution for objects more massive than $10^{14}$ \hMsun as a function of redshift.
The larger amount of massive clusters at higher redshift is a distinctive feature of VDE cosmology.}
\label{img:numberdensity}
\end{center}\end{figure}

Furthermore, if we look at \Fig{img:numberdensity}, where we show the evolution with 
redshift of the number density of objects
above the $M=10^{14}$\hMsun\ threshold, we observe that the most massive structures
in the two cosmologies form at comparable rates. This seems to suggest that in the VDE picture there is a 
subtle balance between the formation of new small haloes and their merging into more massive structures.
Such an effect comes as no surprise if we again take into account that this model has two main opposite, different
features that affect the formation of structures: a strong suppression on all scales induced by the
faster expansion of the universe for a large redshift interval
and an enhancement due to a higher density of matter and a larger power 
on the small scales. 

An interesting consequence of this kind of behaviour 
is that the VDE overabundance of massive objects
may address some recent observational 
tensions of \LCDM; namely, the high redshift of reionization
and the presence of extremely 
massive clusters at $z>1$. 
Recent microwave background observations seem to prefer a high reionization redshift, around
$z\approx10$ combined with a lower normalization of the matter perturbations, $\sigma_8\approx0.8$; whereas
simulations have shown \citep[see, for example,][]{Raicevic:2011} that early reionization can be achieved 
only for $\sigma_8=0.9$ or larger. In VDE, the appearence of dark matter haloes with masses larger than 
$10^{12}$\hMsun\ as early as $z=7$ (while equivalent structures appear in \LCDM\ only for $z>5$) might 
imply also a larger $z_{\rm reion}$, provided the hierarchycal picture of structure formation
holds also in VDE at smaller mass scales. 
On the other hand, the existence of $M>5\times10^{14}$\hMsun\ clusters at $z>1$ \citep[as discussed in ][]{Jee:2009, Brodwin:2010, Foley:2011} has also been considered by many authors 
\citep[e.g.,][]{Hoyle:2010, Baldi:2010, Baldi:2011, Enqvist:2011} as a serious challenge to the standard 
\LCDM\ paradigm; for a more thorough
discussion of this issue in the context of VDE cosmology 
we refer to aforementioned articles as well as \cite{Carlesi:2011}.
However, the comparison to the \LCDM-vde paradigm also shown in \Fig{img:numberdensity} shows that indeed VDE acts
as a source of suppression of structure growth with respect to the enhancement triggered by the increase in 
$\sigma_8$ and $\Omega_M$. This effect is indeed a general result of uncoupled dynamical dark energy models
\citep[]{Grossi:2009, Li:2011b} as the presence of a larger fraction of dark energy at high $z$ enhances Hubble expansion
(as shown in \Fig{img:hratio}) preventing a stronger clustering to take place.

In our case, it is also important to point out that the overprediction of objects at $z=0$
may represent a shortcoming of the model, as observations on the cluster number mass function seem to be in 
contrast with such a prediction (see \cite{Vikhlinin:2009b}, \cite{Wen:2010} and \cite{Burenin:2012}).
Furthermore, we have to keep in mind that these results assume a \LCDM\ fiducial model, while the
use of a different cosmology requires a careful handling of the data and does not allow 
a straightforward comparison to the observations, as they are affected by model-dependent quantities 
like comoving volumes and mass-temperature relations.

\subsection{Void Function}

In order to identify voids, our voidfinder starts with a selection of point
like objects in 3D. These objects can be haloes above a certain mass or a
certain circular velocity or galaxies above a certain luminosity. Thus the
detected voids are characterized by this threshold mass, circular velocity or
luminosity. 
Other voidfinders use different approaches \citep{Colberg:2008}.
The void finding algorithm does not take into account periodic boundary
conditions used in numerical simulations. Therefore, we have periodically
extended the simulation box by 50 \hMpc. In this extended box we represent all
haloes with a mass above the threshold of $5\times10^{12}\hMsun$ as a point. In
this point distribution we search at first the largest empty sphere which is
completely inside the box.  To find the other voids we repeat this procedure
however taking into account the previously found voids. We allow that newly
detected voids intersect with previously detected ones up to 25\% of the
radius of a previously detected larger void.

\begin{figure} 
\includegraphics[angle=270, width=8cm]{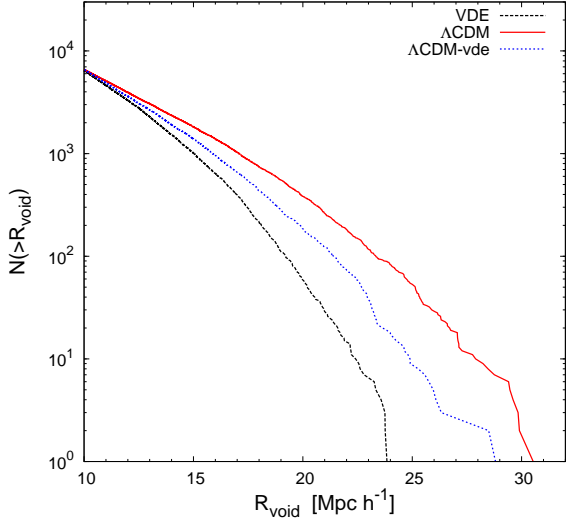}
\caption{
Void function for VDE-0.5, \LCDM-0.5 and \LCDM-vde-0.5 at z=0. For the 
$500$\hMpc\ box we show the cumulative number of empty spheres of radius $R$ which 
do not contain any object with mass larger than $5\times10^{12}$\hMsun.
}
\label{img:voidfunction}
\end{figure}

In \Fig{img:voidfunction} we show the cumulative number of voids with radius larger than
$R_{\rm void}$ the center of which is in the original box.
One can clearly see that for a given void radius there exist more voids in the
{\LCDM} than in th \LCDM-vde and VDE models. The void distribution reflects the
behaviour of the mass function shown in \Fig{img:massfunction}. At redshift $z=0$ there exist
less haloes with $m_h > 5\times10^{12}\hMsun$ in the {\LCDM} model than in the
other two models. Thus on average larger voids are expected.

\subsection{Growth Index}\label{sec:gamma}

\begin{figure}\begin{center}
\includegraphics[angle=270, width=8cm]{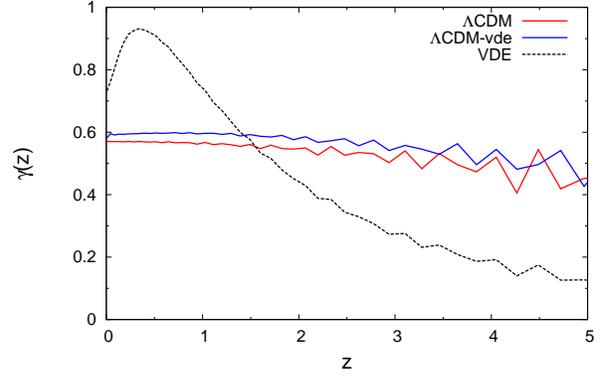}
\caption{Growth index in the VDE and \LCDM\ cosmologies from $z=5$ to $z=0$.
Whereas \LCDM's growth index has an almost constant behaviour with 
a mild dependence on the redshift, VDE changes dramatically from
a regime where growth is relatively suppressed (until $z\approx1.5$) to a relative enhancement
at earlier times, where $\gamma$ becomes larger.}
\label{img:growthindex}
\end{center}\end{figure}

The growth of the perturbations can be related to the evolution of the matter
density parameter by the general relation

\begin{equation}
\Omega_M^{\gamma(a)} = \frac{d \ln(\delta(a))}{d \ln(a)}
\end{equation}

In the standard \LCDM\ cosmology, the exponent $\gamma(a)$  can be approximated by a constant 
value $\gamma \sim 0.55$, although a more detailed calculation shows that this number
is actually redshift dependend \citep[see ][]{Belloso:2011ms}.
In \Fig{img:growthindex} we show the evolution of this growth index $\gamma(z)$ computed 
from our VDE-0.5, \LCDM-vde and \LCDM-0.5 simulations.
As expected, we do observe that in VDE structure formation 
is generally suppressed with respect to \LCDM\ as an effect of the faster expansion rate.
This statement is true until $z\approx 1.5$, when the ratio $H_{VDE}/H_{\LCDM}$ 
starts decreasing causing the steep increase in the growth index, 
eventually reducing again as soon as vector dark energy
enters into the phantom regime (see \Sec{sec:model}), undergoing an accelerated expansion that strongly
suppresses structure formation. This latter change, which takes place at $z\approx0.5$, is
reflected by the peak of $\gamma(z)$, which is reached for the same $z$. 
Actually, as stressed by different parametrizations \citep{Belloso:2011ms}, the growth index is extremely sensitive 
to the value of the equation of state $\omega(z)$, although an explicit form
in terms of VDE cosmology still has to be found.
Indeed, the extremely different behaviour of this parameter at different redshifts 
is an interesting feature that clearly distinguishes the two models in a unique way: 
In fact, parameter-induced modification accounts for a $\approx 5\%$ change for the 
value of the growth factor, as the comparison among \LCDM\ and \LCDM-vde suggests. In this case
we observe a slight increase of the value of $\gamma(z)$ at all redshifts, due to the 
increased growth rate in \LCDM-vde also shown in \Fig{img:growthfactor}. However, these
changes have no impact on the shape of this function, which keeps its mild dependence
on $z$ unaltered.
Therefore $\gamma(z)$ can be effectively used as a tool for model selection, embodying effectively
VDE's peculiar equation of state $\omega(z)$ and expansion history.
Current observational bounds on $\gamma$ constrain only weakly its value at high $z$'s \citep[see e.g. ][]{Nesseris:2008} 
or even favour a higher $\gamma(z=0)$ \citep{Basilakos:2012} in contrast to theoretical calculations based on \LCDM.
In any case, it will surely 
be something to be looked at in the near future, when deep surveys like Euclid \citep{Euclid:2011} will provide
stringent constraints on this quantity \citep{Belloso:2011ms}.

\section{Dark Matter Haloes} \label{sec:results2}
In this section we will discuss properties of (individual) haloes in VDE and
\LCDM. In particular, we will compare the distributions of masses, shape parameter, 
spin parameter, concentrations and formation redshifts as well as the shape of dark matter density profiles.
In this way, we will determine the most important features that characterize on the average 
a single cosmological model. 
But in addition we are also cross correlating haloes in the two models studying
differences on a one-to-one basis. By this we will be able to determine how the properties of a single given structure change 
when switching from one cosmological picture to the other.

\subsection{General properties} \label{sec:general}
To have a reliable description of the general halo properties we need to properly
select our sample from the catalogues, 
in order to include only those objects composed of a number of particle sufficient to 
resolve its internal structure without exceeding statistical uncertainty.
Following \cite{Munoz-Cuartas:2011} and \cite{Prada:2011} we set this number to approximately 500, 
even though other authors (see for example \cite{Maccio:2007} and \cite{Bett:2007}) suggest 
that lower values can be used, too.
However, since we are dealing with different simulations run with particles of different 
mass, the application of this criterion is not straightforward. 
In fact, since our aim is to compare \emph{equivalent} structures
(i.e. structures with the same $M_{200}$) and not structures composed
by an identical number of particles we need to choose our sample imposing a mass threshold 
$M_{th}$.
For the simulations in the $500$\hMpc\ box, we have chosen $M_{th}=5\times10^{13}$\hMsun, which corresponds to haloes formed
by at least 500 particles in VDE and \LCDM-vde and 715 particles in \LCDM; while for the larger $1000$\hMpc\ runs we imposed 
a $M_{th}=3\times10^{14}$\hMsun\ limit, i.e. 380 VDE and \LCDM-vde particles and 545 \LCDM\ ones. 
In the latter set of simulations, we see that we are including also haloes with a 
$\sim 20\%$ less than 500 particles in the VDE and \LCDM-vde cases; this has been done since in the trade off 
between resolution and sample size we have felt more comfortable using a larger number of haloes
at the expense of a slight reduction in accuracy, which will be nonetheless taken into account
when analyzing the results.
The total number of haloes that comply with these conditions in every simulations, as well as 
the number of haloes that satisfy the relaxation criterion which will be discussed in \Sec{sec:relaxation}, 
are shown in \Tab{tab:nhaloes}. The state of virialisation of haloes will only be taken into account
below when investigating the density profiles; for the study of the (distributions of the) two-point correlation functions, the spin,
and even the shape of haloes we prefer to include even un-relaxed objects as they should clearly stick out in the distributions
(if present in large quantities).

\begin{table}
\caption{Number of haloes above the mass (number) threshold $M_{th}$ ($N_{th}$) per simulation. It is also shown
the number of relaxed haloes, defined 
as those complying with the criterion introduced in \Sec{sec:relaxation}.
}
\begin{center}
\begin{tabular}{ccccc}
\hline
Simulation & $M_{th}$ & $N_{th}$  & $N$ total & $N$ relaxed  \\
\hline
\LCDM-0.5 & $5\times10^{13}$\hMsun & 715 & 1704  & 1370  \\
\LCDM-vde-0.5 &$5\times10^{13}$\hMsun & 500 & 5898 & 5220 \\ 
VDE-0.5 &  $5\times10^{13}$\hMsun & 500 & 6274 & 5569 \\
\LCDM-1 & $3\times10^{14}$\hMsun & 545 & 4045 & 3533 \\
\LCDM-vde-1 & $3\times10^{14}$\hMsun & 380 & 9072 & 8117 \\
VDE-1 &  $3\times10^{14}$\hMsun & 380 & 12174 & 11508 \\
\hline
\end{tabular}
\label{tab:nhaloes}
\end{center}
\end{table}


\subsubsection{Correlation function}

\begin{figure}\begin{center}
\includegraphics[angle=270, width=8cm]{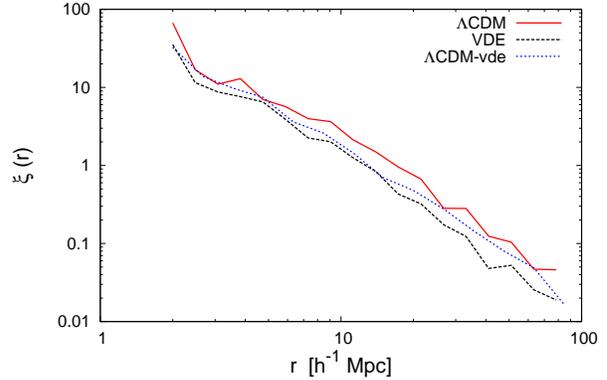}
\caption{The two point correlation function for objects in the $500$\hMpc\ simulations more massive than $5\times10^{13}$\hMsun.
}\label{img:correlation}
\end{center}\end{figure}

To study the clustering properties of the haloes in VDE cosmology, 
 we computed the two point correlation 
function using the definition:

\beq
\xi(r) = \frac{V}{N^2} \sum_{i=1}^{N} \frac{n_i(r;\Delta r)}{v(r; \Delta r)} - 1
\eeq

\noindent
where $N$ is the total number of objects above the given mass threshold 
in the simulation volume $V$, and $n_i$ is 
the total number of objects within a shell of volume $v$ and thickness $\Delta r$
(of constant logarithmic spacing in $r$) centered at the $i$th object.
In this case, we have limited our analysis to the 500\hMpc\ boxes, ignoring the 1\hGpc\
due to their lack of small scale resolution.
The results are plotted in \Fig{img:correlation}, where we can see that the $\xi(r)$ 
is slightly smaller at all scales in VDE.
Although in principle we would expect VDE cosmology to have an enhanced clustering pattern 
due to the increased distribution of massive objects observed in the mass function, the
$N^{-2}$ dependence of the two point correlation function drags the total value down, 
making the final distribution function smaller than in \LCDM.
In fact, a similar behaviour can be observed for \LCDM-vde; with a two point correlation
function below \LCDM\ at practically all scales.
In \Tab{tab:cbestfit} 
we show the results of fitting $\xi(r)$ to a power law $(r_0/r)^{\gamma}$ from which we see that
VDE is characterized by a smaller correlation length $r_0$ and a steeper slope $\gamma$.


\subsubsection{Spin parameter, shape and triaxiality}
\begin{figure}
\begin{center}
$\begin{array}{cc}
\includegraphics[angle=270, width=8cm]{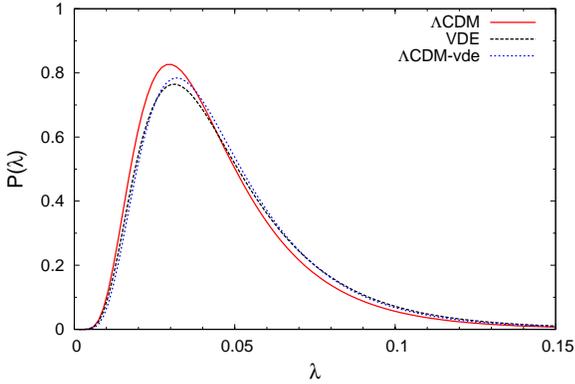}
\end{array}$
\caption{Spin parameter versus the analytical log norm distribution calculated 
with the best fit parameters. The fit has been performed using the combined sample of
haloes above $5\times10^{13}$\hMsun\ belonging to the three 500 \hMpc\ boxes with
those above the $3\times10^{14}$\hMsun limit in the 1\hGpc\ boxes. 
}
\label{img:spin}
\end{center}\end{figure}

Rotational properties of the haloes can be studied using the so called spin parameter $\lambda$, 
a dimensionless number that measures the degree of rotational support of the halo.
Following \cite{Bullock:2001}, we define it as

\begin{equation}
\lambda = \frac{L_{200}}{\sqrt{2}M_{200}V_{200}R_{200}}
\end{equation}

\noindent
where the quantities $L$ (the total angular momentum), $M$ (total mass), $V$ (circular velocity) 
and $R$ (radius) are all taken at the point where the average halo density becomes 200
times the critical density.
Different authors have found \citep[e.g.][]{Barnes:1987, Warren:1992, Cole:1996, Gardner:2001, Bullock:2001, Maccio:2007, Maccio:2008, Munoz-Cuartas:2011}
that the distribution of this parameter is of lognormal type

\begin{equation}
P(\lambda) = \frac{1}{\lambda\sigma_0^2\sqrt{2\pi}} \exp \left[ - \frac{\ln^2(\lambda/\lambda_0)}{2\sigma_0^2}\right] \ ,
\end{equation}

\noindent
even though there are recent claims that this distribution has to be slightly modified \citep{Bett:2007}.

Fitting the above function to our numerical sample by a non-linear Levenberg-Marquardt 
least square fit we find a remarkably good agreement, shown in \Fig{img:spin} for the combined set of haloes
of the 500\hMpc\ and 1\hGpc\ simulations.
It is clear that the three models present
no substantial difference in the values of these distributions, meaning that the change of cosmology
has no impact on the rotational support of the dark matter structures.
\\

The shape of three dimensional haloes can be modelled
as an ellipsoidal distribution of particles \citep{Jing:2002, Allgood:2006}, 
characterized by the three axis 
$a\geq b\geq c$
computed by \texttt{AHF} as the eigenvalues of the inertia tensor 

\beq
I_{i,j} = \sum_n x_{i,n} x_{j,n}
\eeq

\noindent
which is in turn obtained summing over all the coordinates of the particles belonging to the halo.

We define the shape parameter $s$ and the triaxiality parameter $T$ as

\beq
s = \frac{c}{a} \qquad T = \frac{a^2 - b^2}{a^2 - c^2}
\eeq

\noindent
and we calculate the probability distributions $P(T)$ and $P(s)$ 
of the above parameters for all the objects above the aforementioned mass thresholds in our cosmological simulations, 
to see whether the VDE picture of structure formation induces changes in the average shape and triaxiality.
Similarly to the previous case, we found again that halo shapes and triaxialities 
remain practically unaltered by VDE cosmolgy.
This result could be expected, keeping in mind that VDE only affect background evolution: Once that 
structures start to form, detaching from the background evolution, they become affected by gravitational attraction only.
Therefore, the internal structure of dark matter haloes remains generally unaltered by the presence of 
an uninteracting form of dark energy and cannot be used to discriminate between alternative cosmological
paradigms. We have also verified that these results hold also when taking into account 
different halo samples separately, i.e. the massive ones of the 1\hGpc\ simulations
and the smaller belonging to the 500\hMpc\ boxes.


\subsubsection{Unrelaxed haloes}
\label{sec:relaxation}
Before moving to the discussion of the properties of internal structure of the haloes, and in particular the density profile,
we need to introduce and motivate a second criterion of selection for our halo sample, related to the
degree of \emph{relaxation} of the halo. 
An additional check is necessary since only a fraction of the structures identified 
in our catalogues completely satisfies the virial 
condition. In unvirialized structures, infalling matter and merger phenomena may occour,
heavily affecting the halo shape and thus making 
the determination of radial density profiles and concentrations unreliable.
In fact, unrelaxed haloes are most likely to differ 
from an idealized spherical or ellipsoidal shape since they have
a highly asymmetric matter distribution, which in turn makes the determination of the halo center an ill-defined problem, 
as discussed by \cite{Maccio:2007} and \cite{Munoz-Cuartas:2011}.
Our halo finder \texttt{AHF} does not
directly discriminate between virialized and unvirialized structures giving catalogues containing both types of objects; however, it provides
kinetic $K$ and potential energy $U$ for every halo identified, thus making 
the computation of the viral ratio $2K/|U|$ straightforward.
Following one of the criteria used by \cite{Prada:2011}, 
we will consider as relaxed all the haloes satisfying the condition

\beq
\frac{2K}{|U|} - 1 < 0.5
\label{eq:vircond}
\eeq

\noindent
without introducing additional parameters.
Alternative ways of identifying unrelaxed structures can be found throughout the literature \citep[e.g.][]{Maccio:2007, Bett:2007, Neto:2007, Knebe:2008, Prada:2011,
Munoz-Cuartas:2011,Power:2011}; but since the results they give are qualitatively similar for 
reasons of computational speed and simplicity we will not make use of them.
The total number of haloes satisfying the relaxation condition 
is shown for every cosmology in \Tab{tab:nhaloes}.

\subsubsection{Density profiles} \label{sec:nfw}
$N$-body simulations have shown that dark matter haloes can be
described by a Navarro Frenk White (NFW) profile \citep{NFW:1996}, 
which is given by

\beq
\rho(r) = \frac{\rho_0}{\frac{r}{r_s}(1+\frac{r}{r_s})^{2}}
\label{eq:nfw}
\eeq

\noindent
where the $r_s$, the so called scale radius, and the $\rho_0$ are in principle two 
free parameters that depend on the particular halo structure.
But $\rho_0$ can be written as a function of the critical density as $\rho_0=\delta_c\rho_c$,
where 

$$
\delta_c = \frac{200}{3}\frac{c^3}{\log(1+c)-\frac{c}{1+c}}
$$

\noindent
and  $c=r_{\rm vir}/r_s$ is the \emph{concentration} of the halo relating the virial radius $r_v$($=r_{200}$ in our case) to the scale radius $r_s$, which will be discussed in detail in the 
following subsection. This description is generally valid for \LCDM, but simulations of ever increased
resolution have actually revealed that the very central regions are not following the slope advocated
by the NFW formula but rather follow a Sersic or Einasto profile \citep[cf. ][]{Navarro:2004,Stadel:2009}.

Here we want to check to which degree the modified cosmological background affects 
the distribution of matter inside dark matter haloes, i.e. its density profile. 
All our (relaxed) objects in all simulations have been fitted to the \Eq{eq:nfw}, and to estimate the goodness of this fit we compute for each halo its corresponding $\chi^2$, defined in the usual way

\beq
\chi^2 = \sum_i \frac{(\rho_i^{\rm (th)}-\rho_i^{\rm (num)})^2}{\Delta\rho_i^{\rm(num)}}
\eeq

\noindent
where the $\rho_i$'s are the numerical and theoretical overdensities in units of the critical density $\rho_c$ 
at the $i$-th radial bin and  $\Delta\rho_i$ is the numerical Poissonian error on the numerical estimate.
Since different halo profiles will be in general described by a different number of radial bins\footnote{Note that our halo finder \texttt{AHF} uses logarithmically spaced radial bins whose number depends on the halo mass, i.e. more massive haloes will be covered with more bins.}, to
make our comparison between different simulations and haloes consistent we need to use the reduced $\chi^2$

\beq
\chi^2_{\rm red} = \frac{\chi^2}{N_{pts}-N_{dof}-1}
\eeq

\noindent
where $N_{pts}$ is the total number of points used (i.e. total number of radial bins) and $N_{dof}$ is the number of degrees of freedom (free parameters).

\begin{figure*}
\begin{center}
$\begin{array}{cc}
\includegraphics[angle=270, width=8cm]{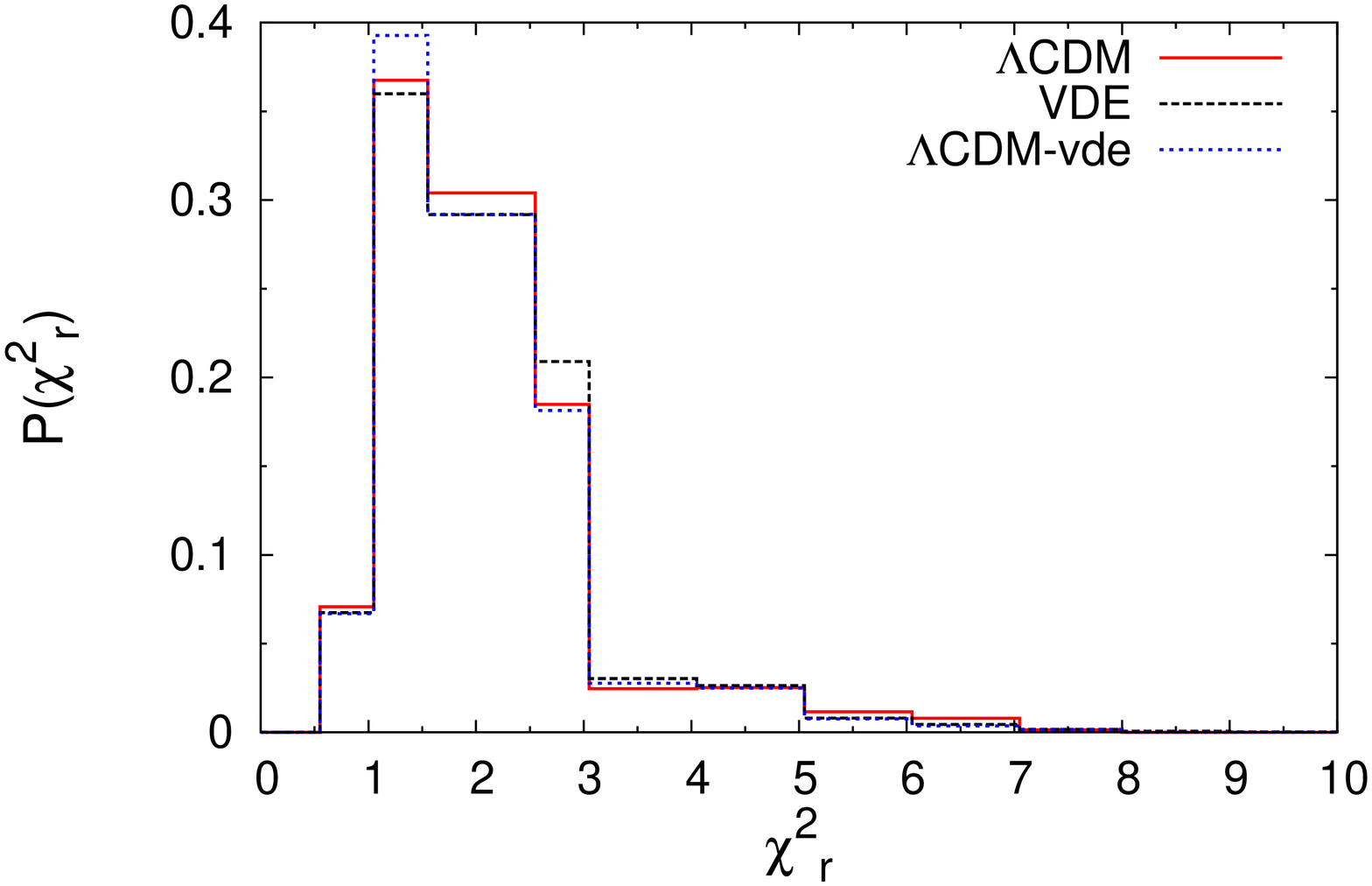}
\includegraphics[angle=270, width=8cm]{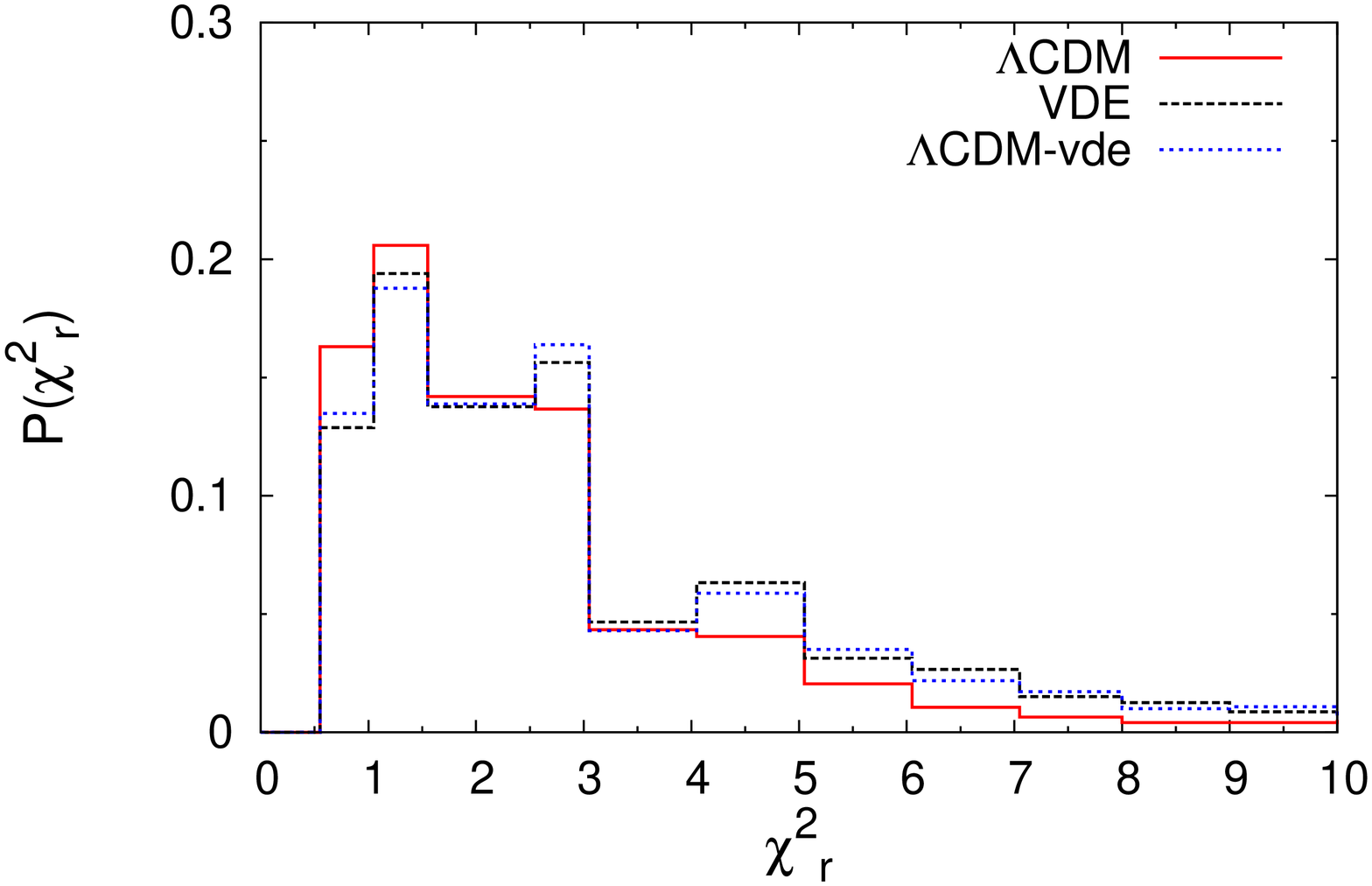}
\end{array}$
\caption{Reduced $\chi^2$ distribution for the best fit to a NFW profile, on the vertical axis we plot 
the total fraction of haloes whose reduced $\chi^2$ falls within the horizontal axis bin value.
This is shown for relaxed haloes above the $5\times10^{13}$\hMsun\ 
threshold belonging to the VDE-0.5, \LCDM-vde-0.5 and the \LCDM-0.5 simulations (left panel) 
as well as for those above the $3\times10^{14}$\hMsun\ threshold belonging to VDE-1, \LCDM-vde-1 
and \LCDM-1 (right panel).
The distributions show no particular difference among the three cosmologies; however, in the three 1\hGpc\ 
simulations we notice how lower resolution affects the $\chi^2$ distribution, resulting in a ticker tail
at higher values compared to the 500\hMpc\ case, meaning that the fit to a NFW is on the average worse.
}
\label{img:nfwfit}
\end{center}\end{figure*}

The comparison of the distributions of the reduced $\chi^2$ values for \LCDM-vde, \LCDM\ and VDE haloes belonging
to the two set of 500\hMpc\ and 1\hGpc\ simulations, 
shown in \Fig{img:nfwfit},
allows us to determine again that no substantial 
difference is induced by the VDE picture, 
for the same reasons discussed in the case of spin, shape and triaxiality distributions.
The standard description of dark matter structures is thus not affected by the presence of a VDE.

\subsubsection{Halo concentrations}
In the last step of the analysis of the general properties of haloes we will turn to 
concentrations, which characterize the halo inner density compared to the outer part.
This parameter is usually defined as

\beq
c = \frac{r_{\rm vir}}{r_s}
\eeq

\noindent
where $r_s$ is the previously introduced scale radius, obtained through the best fit 
procedure of the density distribution to a NFW profile. 
We would like to remind that concentrations are correlated to the formation time of the halo, since structures that
collapsed earlier tend to have a more compact center due to the fact that it has
more time to accrete matter from the outer parts. 
Dynamical dark energy cosmologies  generically imply larger $c$ values
as a consequence of earlier structure formation, as found in works like those by \cite{Dolag:2004}, 
\cite{Bartelmann:2006} and \cite{Grossi:2009}. In fact, since the presence of early dark energy
usually suppresses structure growth, in order to reproduce current 
observations we need to trigger an earlier start of the formation process, which on  average
yields a higher value for the halo concentrations.
However, this result does not hold in the case of coupled dark energy, where the increased 
clustering strength induced by a fifth force sets a later start of structure formation, 
as discussed in \cite{Baldi:2010a}.

\begin{figure}\begin{center}
$\begin{array}{cc}
\includegraphics[angle=270,width=8cm]{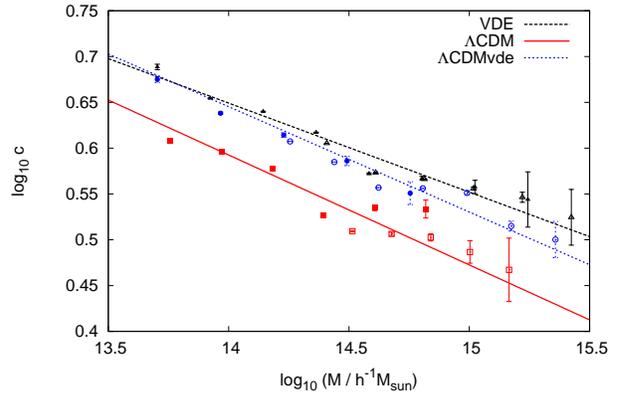}
\end{array}$
\caption{Best fit of the mass-concentration relation for the combined sample of relaxed haloes 
belonging to all the \LCDM, \LCDM-vde and VDE simulations. The points represent the average concentration
values for the relaxed haloes in the corresponding mass bin; circles are for \LCDM, triangles for \LCDM-vde and 
squares for VDE. Empty dots stand for bins determined using haloes
 belonging to the 1 \hGpc\ simulations; filled ones refer to the 500 \hMpc\ ones.
The Poissonian error bars are computed using the number 
of selected haloes within each mass bin.
}
\label{img:c-pl-fit}
\end{center}\end{figure}

In the hierarchical picture of structure formation, concentrations are usually inversely
correlated to the halo mass as more massive objects form later; $N$-body simulations 
\citep{Dolag:2004,Prada:2011,Munoz-Cuartas:2011} and observations
\citep{Comerford:2007,Okabe:2010,Sereno:2011} have in fact shown 
that the relation between the two quantities can be written as a power law of the form 

\beq
\log{c} = a(z)\log(\frac{M_{200}}{\hMsun}) + b(z)
\label{eq:mass-c}
\eeq

\noindent
where $a(z)$ and $b(z)$ can have explicit parametrizations as functions of redshift and cosmology
\citep[see e.g.][]{Neto:2007,Prada:2011,Munoz-Cuartas:2011}.
We can use our selected halo samples at $z=0$ from the 500 \hMpc\ and 1 \hGpc\ simulations 
to obtain the $a(z=0)$ and $b(z=0)$ values for the \LCDM, \LCDM-vde and VDE 
cosmologies; the results of the best fit procedure to \Eq{eq:mass-c} are shown in \Tab{tab:cbestfit}.

These values are in good agreement with the ones found by, for instance, \cite{Dolag:2004}, \cite{Maccio:2008} and 
\cite{Munoz-Cuartas:2011} (who quote for \LCDM\ values of $a(z=0)\approx-0.097$ and $b\approx2.01$); 
the $\sim10\%$ discrepancy observed with their results is due to the fact that our results
are obtained over a smaller mass range, $5\times10^{13}$--$2\times10^{15}$ \hMsun, whereas the 
previously cited works study it over an interval larger by more than three orders of magnitude, 
$10^{10}$--$10^{15}$\hMsun.
Still, according to our results, the $c-M$ relation for both the VDE and \LCDM-vde case
is characterized by a shallower $a$ exponent and a larger $b$. Although the magnitude of these 
changes is different in the two models, we can safely conclude that also in this case the 
results are mainly parameter-driven, i.e. due to the larger 
value of $\Omega_{M}$. Furthermore, the large error bars for $M>10^{15}$ \hMsun\
scales, due to the low statistics of massive haloes complying the relaxation requirements, makes it
difficult to determine to which extent the differences in the best fit relations among 
\LCDM-vde and VDE could be eventually reduced in the presence of a larger sample.

\begin{table}
\caption{Best fit values for the mass-concentration relation for $z=0$, obtained fitting the relation 
\Eq{eq:mass-c} to the relaxed haloes concentrations and the two point correlation 
function to a power law $ (r_0 / r)^{\gamma}$ for the \LCDM, \LCDM-vde and VDE cosmologies.
$r_0$ values are given in \hMpc.
}
\begin{center}
\begin{tabular}{ccccc}
\hline
Model & $a$ & $b$ & $r_0$ & $\gamma$ \\
\hline
\LCDM & -0.115  & 2.11 & 13.4 & -1.79\\
\LCDM-vde & -0.112  & 2.21 & 12.1 & -1.91 \\
VDE & -0.098 & 2.17 & 10.1 & -1.94\\
\hline
\end{tabular}
\label{tab:cbestfit}
\end{center}
\end{table}

\begin{figure}\begin{center}
$\begin{array}{cc}
\includegraphics[angle=270,width=8cm]{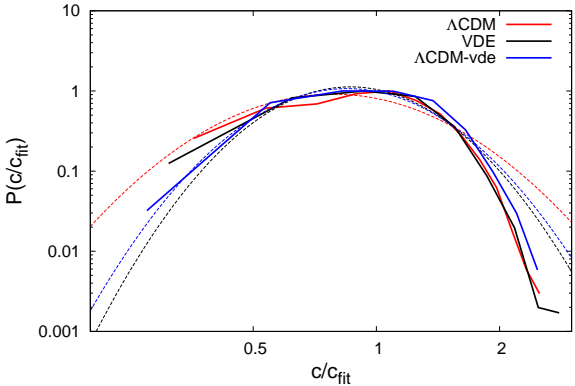}
\end{array}$
\caption{Distribution of the ratio between the actual concentration and the expected one (cf. \Eq{eq:mass-c}) and its fit to a lognormal distribution. }
\label{img:concentration}
\end{center}\end{figure}

We also need to mention that in our simulations the actual halo concentrations do not precisely follow equation
\Eq{eq:mass-c} but rather scatter around it, as can be seen in \Fig{img:c-pl-fit}, where the 
average $c$ per mass bin is plotted against the corresponding best fit relations.
This is not really surprising, since observations \citep{Sereno:2011} and $N$-body simulations \citep{Dolag:2004} have 
shown that halo concentrations are lognormally distributed 
around their theoretical value calculated using \Eq{eq:mass-c}.
In \Fig{img:concentration} we show that this is indeed the case: the distribution of the $c(M)/c_{fit}(M)$, 
where $c_{fit}(M)$ is the theoretical concentration value for a halo of mass $M$,
extremely close to a lognormal one with an almost model independent dispersion $\sigma \approx 0.4$.


\subsection{Cross Correlation}
The next step in our analysis consists of studying the properties of the (most massive) 
cross correlated objects found in the three models at $z=0$. Whereas in the previous section
our focus was on the distribution of halo properties, this time we aim 
at understanding how they change switching from one model to another. 

The identification of "sister haloes" among the different cosmologies can be done 
using the \texttt{AHF} tool \texttt{MergerTree}, which determines correlated structures by matching 
individual particles IDs in different simulation snapshots. For a more elaborate discussion of its mode of operation
we refer the reader to Section~2.4 in \cite{Libeskind:2010} where it has been described in greater detail.
This time we decided to restrict our halo sample further only picking the first 1000 most massive (\LCDM) haloes. 
The criterion of halo relaxation has of course also been taken into account when dealing 
with profiles and concentrations.


\subsubsection{Mass and spin parameter} 

\begin{figure*}\begin{center}
\includegraphics[angle=270, width=17cm]{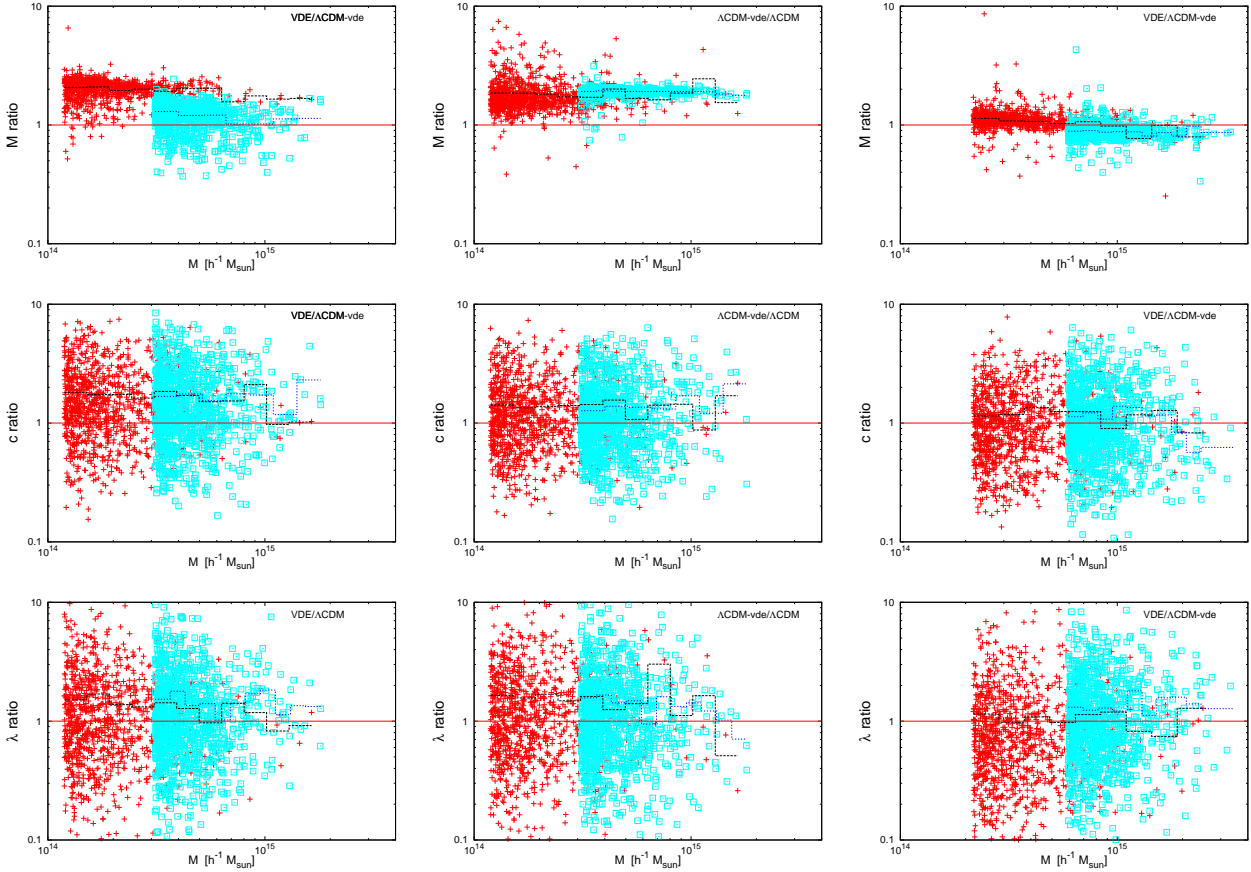}
\caption{Mass and spin parameter correlation ratios for the first 1000 (relaxed and possibly unrelaxed) haloes.
Panels on the left show the VDE/\LCDM\ results, the ones in the center to \LCDM-vde/\LCDM while on the right
the ratio VDE/\LCDM-vde is plotted. 
Cross identified objects are characterized by larger masses in VDE and \LCDM-vde as
a consequence of the higher $\Omega_{M}$ and $\sigma_8$ normalization value.}
\label{img:crosscorr}
\end{center}\end{figure*}

In the two upper panels of \Fig{img:crosscorr} we show the ratios of the masses $M$ and spin parameter $\lambda$ 
for all the cross correlated sets of simulations; in each panel we show
the ratios for the 500\hMpc\ simulation boxes while on the right the 1\hGpc\ ones.
Both VDE and \LCDM-vde show average mass and spin values scattered around values
larger than one when compared to \LCDM, whereas the cross comparison of VDE to \LCDM-vde 
shows average ratios close to unity at all mass scales.
This substantial increase in the ratios is due
to the earlier beginning of structure formation, triggered by the larger $\Omega_M$ and $\sigma_8$,
as the comparison VDE/\LCDM-vde shows.
As we already did in \Sec{sec:general} when looking at the halo properties in general, we conclude 
that also when observing the same halo evolved under different cosmologies, the main 
effects are determined exclusively by the set of cosmological parameters chosen, being the
imprint of the cosmological background evolution substantially negligible in this case.
This makes the identification of a cosmic vector through
the determination of halo properties impossible, since the
background dynamics, which distinguishes VDE 
from any other non interacting dynamical dark energy model,
does not leave any observable imprint on these scales.


\subsubsection{Halo concentrations and internal structure}

As done in the previous section, in the determination of halo profiles and concentrations properties we  
 discard unrelaxed haloes, but this time in a way so that our halo sample will still be composed of the 
first 1000 haloes satisfying condition \Eq{eq:vircond}. This same halo sample has been used also 
in the study of the $M_{vir}-z_{\rm form}$, in order to be able to compare these results with the ones
obtained from concentrations consistently -- although in principle formation redshifts are well 
defined even for unrelaxed haloes.
Again, our procedure consists in fitting all the selected structure to a NFW profile, 
from which we will be able to derive the concentration parameter $c$ and a measure for the quality of the fit $\chi^2$; we will then compare these 
results in each cross identified objects to see how a given halo structure changes 
when evolved under a different cosmology.
Although not shown here, no particular trend in the differences among \LCDM, \LCDM-vde and VDE 
pictures has been be found for either NFW $\chi^2$, shape and triaxiality; since 
in all the cases the ratios of these properties among cross correlated haloes are 
centered around unity.
Not surprisingly, we also find again a generally higher average value for the concentrations 
in VDE and \LCDM-vde with respect to \LCDM, (see \Fig{img:crosscorr}) a result 
which again can be explained by the larger value of $\Omega_M$ 
and $\sigma_8$. Similar concentrations for VDE and \LCDM-vde haloes, shown in the upper right panel of
\Fig{img:crosscorr} can be also understood as a consequences of the similar masses 
of the haloes examined and the similar $c-M$ relations found for the two cosmologies.
However, even if from \LCDM-vde cosmology we conclude that the different choice of $\Omega_{M}$,
can explain in this case higher halo concentration, we need to remind that such a results is 
also a general feature of the dynamical nature
of the dark energy fluid, as already found by \cite{Dolag:2004},
\cite{Bartelmann:2006} and \cite{Grossi:2009}.

\section{Conclusions} \label{sec:conclusions}
In this work we presented an in-depth analysis of the results of a 
series of $N$-body dark matter only simulations of the Vector Dark Energy cosmology proposed by \cite{BeltranMaroto:2008}.
The main emphasis has been on the comparison to the standard \LCDM\ paradigm, using 
a mirror simulation with identical number of particles, random seed for the initial conditions, box size and starting redshift. 
An additional series of simulations for a \LCDM-vde cosmology have also been run 
using the VDE values for $\Omega_M$ and $\sigma_8$ within a standard \LCDM\ picture,
to disentangle the effects of the parameter induced modifications 
to the dynamical ones coming directly from the VDE model.

The use of a modified version of the \texttt{GADGET-2} code required us to check the 
results with particular care.
A consistency check of our simulations was performed by comparing the numerical results for the evolution of the growth factor
to the analytical calculations, finding an excellent agreement between the two. 
We further had to adapt the halo finding procedure, due to the fact that the critical density
as a function of redshift $\rho_c(z)$, entering
the definition of the halo edges, takes different values in VDE.
Once halo catalogues had been obtained, we carried our analysis at two different levels, namely:

\begin{itemize}
\item we studied the very large scale clustering pattern through the computation of 
matter power spectra, mass,  void, and two-point correlation functions;
\item we analyzed halo structure, comparing statistical distributions and averages of 
spin parameters, concentrations, masses and shapes.
\end{itemize}

In the first point, making use of the full set of simulations, 
our analysis covered the whole masse range $10^{12}$--$10^{15}$ \hMsun\ as well as
different redshifts, so that we could make specific VDE model predictions 
for the number density evolution $n(>M,z)$ and growth index $\gamma(z)$.
A distinctive behaviour, very far from the standard \LCDM\ results, 
has been found for  $\gamma(z)$, and, in particular, 
for the mass function that in VDE cosmology can be up to 10 times larger than the standard \LCDM\ one.
The latter result is due to the earlier onset of structure formation and we have mentioned
how it can be used to address current \LCDM\ observational tensions with 
large clusters at $z>1$ and possibly with early reionization epoch \citep[cf. also][]{Carlesi:2011}.

Computing the cumulative mass function at different redshifts and making use of the \LCDM-vde simulations 
we have also observed how the condition $H_{VDE}(z) > H_{\LCDM}(z)$, holding up to
$z \approx 1$, induces a relative suppression of structure growth in this cosmological model, 
an effect that clashes with the increased matter density and $\sigma_8$.
In fact, while on the one hand higher values of these parameters enhance the formation of a larger number
of objects, on the other hand, background dynamics suppresses clustering and growth. 
The interplay and  relative size of these effects has been studied using the \LCDM-vde simulations, 
showing that, for example, faster expansion in the past determines for VDE an expectation of clusters 
with $M>10^{14}$\hMsun\ up to $\approx 5$ times
smaller than what a simple increase in $\sigma_8$ and $\Omega_M$ would determine.
This effect has been also seen in the void distribution, where 
suppression of clustering prevents small structures to merge into larger one and to rather spread in the field,
so that underdense regions happen to be smaller and rarer than 
in \LCDM\ and \LCDM-vde. In these latter cosmologies, in fact, 
a higher contrast between populated and less populated regions is observed both in the power spectrum and 
in the colour coded matter density.

In the second part of our work we have focused on the study of internal halo structure.
We found that VDE cosmology does not induce deviations in the functional form of the dark matter halo density profiles, which 
are still well described by a NFW \citep{NFW:1996} profile, nor in the distributions 
for the concentrations and spin parameters,
which are of the lognormal type as in \LCDM.
Shape and triaxiality are also unaffected: the distributions for the relative parameters are identical
and peaked in at the same values in all the three cosmologies. 
The above results are a direct consequence of the fact that dark matter haloes, once detached from the 
general background evolution driven by the cosmic vector, evolve by means of gravitational attraction 
only; which is unaffected by the specific nature of dark energy. 
A net effect can be seen in masses, whose average values 
tend to be on the larger than in the \LCDM\ case by a factor of $\approx 2$,
a straightforward consequence of the larger $\Omega_M$ and $\sigma_8$, 
as can be shown by a direct comparison of VDE to \LCDM-vde results, that turn out extremely close in these cases.
On the other hand, the different background evolution seems to affect $c-M$ relations only slightly, 
changing the power law index $a(z)$ and normalization $b(z)$ by a $15\%$. 
In this case we have also found that these values in general agree with previous results from early dark energy
studies such as those by \cite{Dolag:2004}, even though in this case it would certainly be
necessary to test the relation down to smaller mass scales, where a better tuning of the parameter would be also possible, and with a larger statistics on the higher scales.
However, in general, most of the halo-level effects which seem to characterize VDE can 
be simply explained in terms of the different cosmological parameters, as we did comparing 
these results to the outcomes of \LCDM-vde simulations.
For the first time then, through the results of the series of $N$-body simulations, we have shown that
VDE cosmology provides a viable environment for structure formation, also alleviating some observational tensions
emerging with \LCDM.
We have seen how the peculiar dynamics of this model leaves its imprint on structure formation and growth, and 
in particular, how it affects predictions for large scale clustering and halo properties. 
However, a close comparison of the deep non-linear regime results with different sets of observational data
still needs to be performed, challenging us to improve the accuracy of our simulations and at the same 
time devise new and reliable tests which may shed some light not only on VDE but on the nature of dark energy in general.

\section*{Acknowledgements}
We would like to thank Juan Garc\'ia-Bellido for his interesting suggestions and discussions.
EC is supported by the MareNostrum project funded by the Spanish
Ministerio de Ciencia e Innovacion (MICINN) under grant
no. AYA2009-13875-C03-02 and MultiDark Consolider project under grant
CSD2009-00064.  
EC also acknowledges partial support from the European 
Union FP7 ITN INVISIBLES (Marie Curie Actions, PITN- GA-2011- 289442).
AK acknowledges support by the MICINN's Ramon y Cajal
programme as well as the grants AYA 2009-13875-C03-02,
AYA2009-12792-C03-03, CSD2009-00064, and CAM S2009/ESP-1496.
G. Yepes would like to thank the MICINN for financial support under 
grants  AYA 2009-13875-C03, FPA 2009-08958, and the SyeC Consolider 
project CSD2007-00050.
JBJ is supported by the
Ministerio de Educaci\'on under the postdoctoral contract EX2009-0305
and also wishes to acknowledge support from the Norwegian Research
Council under the YGGDRASIL programme 2009-2010 and the NILS mobility
project grant UCM-EEA-ABEL-03-2010.  We also acknowledge support from
MICINN (Spain) project numbers FIS 2008-01323, FPA 2008-00592, CAM/UCM 910309 and FIS2011-23000. 
The simulations used in this work were performed  in the Marenostrum 
supercomputer  at Barcelona Supercomputing Center (BSC).

\bibliographystyle{mn2e}
\bibliography{biblio}

\bsp

\label{lastpage}

\end{document}